\documentclass[pdflatex,sn-mathphys-num, iicol]{sn-jnl}

\usepackage{graphicx}%
\usepackage{multirow}%
\usepackage{amsmath,amssymb,amsfonts}%
\usepackage{amsthm}%
\usepackage{mathrsfs}%
\usepackage[title]{appendix}%
\usepackage{xcolor}%
\usepackage{textcomp}%
\usepackage{manyfoot}%
\usepackage{booktabs}%
\usepackage{tabulary}
\newcolumntype{K}[1]{>{\centering\arraybackslash}p{#1}}
\usepackage{diagbox}
\definecolor{deepskyblue}{RGB}{0, 191, 255}
\definecolor{nrichpink}{RGB}{231, 140, 198}
\usepackage{rotating}
\usepackage{subcaption}
\usepackage{extdash}
\usepackage{float}

\begin{document}

\title{Systematic calculations of $\beta$-delayed fission probabilities using a microscopic framework}

\author*[1,2,3]{\fnm{S.} \sur{Bara}}\email{bara@lpccaen.in2p3.fr}

\author[1]{\fnm{T.~E.} \sur{Cocolios}}

\author[4,5]{\fnm{S.} \sur{Goriely}}
\author[4,5]{\fnm{W.} \sur{Ryssens}}
\author[4,5]{\fnm{A.} \sur{Sánchez-Fernández}}
\author[1]{\fnm{P.} \sur{Van~Duppen}}

\affil[1]{\orgdiv{Instituut voor Kern- en Stralingsfysica}, \orgname{KU Leuven}, \orgaddress{\city{Leuven}, \country{Belgium}}}
\affil*[2]{\orgdiv{Université de Caen Normandie}, \orgname{ENSICAEN, CNRS/IN2P3, LPC Caen UMR6534}, \orgaddress{\city{Caen}, \country{France}}}
\affil*[3]{\orgname{Grand Acc\'el\'erateur National d’Ions Lourds}, \orgaddress{\city{Caen}, \country{France}}}
\affil[4]{\orgdiv{Institut d'Astronomie et d'Astrophysique}, \orgname{Université Libre de Bruxelles}, \orgaddress{\city{Brussels}, \country{Belgium}}}
\affil[5]{\orgdiv{Brussels Laboratory of the Universe – BLU-ULB}, \orgaddress{\city{Brussels}, \country{Belgium}}}


\abstract{The process of $\beta$-delayed fission ($\beta$DF) consists of a mother nucleus $\beta$-decaying to different states of a daughter isotope, followed by fission of the latter with a probability that strongly depends on the energy of the states populated in the $\beta$ decay. Studying this process experimentally can be challenging due to the typically small $\beta$DF probabilities. The theoretical framework presented in this work was developed to globally characterise $\beta$DF probabilities. This approach uses energy density functional based models to obtain the necessary inputs for the final probabilities, which are obtained through the nuclear reaction code TALYS. Calculations were performed for all isotopes with an experimentally measured value of $\beta$DF probabilities. Following the optimisation of the theoretical framework to better reproduce known data, a global RMS deviation of about two orders of magnitude was achieved, an accuracy that can be used to identify promising cases for future experimental $\beta$DF studies. 
}

 
\maketitle

\section{Introduction}\label{introduction}
Beta-delayed fission ($\beta$DF) is a two-step process that begins with the $\beta$ decay of a mother nucleus and ends with the fission of the daughter, provided the energy of the states populated by the mother $\beta$ decay is sufficiently high with respect to the fission barrier of the daughter~\cite{Bara2025}. In this process, the excitation energy of the fissioning daughter nucleus is limited by the Q$_\beta$ value of the mother, which is typically smaller than the primary barrier of the daughter. Hence, $\beta$DF can be classified as low-energy fission. This makes it of particular interest from the point of view of nuclear structure as the process will be governed by the microscopic properties of the nuclei involved. Moreover, $\beta$DF is of interest for nuclear astrophysics due to the role it might play in the termination of the r-process nucleosynthesis~\cite{Goriely2015,Cowan2021}. This nucleosynthesis pathway is governed by a competition between neutron capture that creates heavier nuclei, and $\beta^-$ decay, which increases the proton number (Z) and brings the produced nuclei towards stability. When the r-process reaches the region of heavy neutron-rich nuclei, fission begins to become one of the main decay channels, leading to the termination of the r-process and, as a consequence, to the redistribution of the nuclear abundances. The resulting fission fragments (FF) return to lighter regions of the nuclear chart, preventing the formation of heavier systems. Different types of fission can play a role in the termination of the r-process, including $\beta$DF~\cite{Goriely2015}. Therefore, it is important to determine $\beta$DF probabilities ($\text{P}_{\beta DF}$) for the nuclei involved in the r-process, which are, however, too heavy and too neutron-rich to be produced in laboratories. Consequently, models must be developed to obtain theoretical estimates of $\beta$DF probabilities. These models would therefore benefit from experimental inputs as close as possible to the region of interest for the r-process calculations, which involve highly exotic nuclei. 

Since the nuclei in this region are more difficult to study experimentally, the more accessible cases from the neutron-deficient side of the nuclear chart can also be used to test and improve the reliability and accuracy of the models. The majority of $\beta$DF cases studied so far are found in the neutron-deficient lead or trans-uranium regions of the nuclear chart, while only very few neutron-rich cases have been investigated up to now. This makes the benchmarking of models for $\beta$DF calculations difficult in the neutron-rich region. Experimentally, studies of $\beta$DF remain challenging: an experiment at ISOLDE (CERN) dedicated to the search for $\beta$DF in $^{230,232}$Fr and $^{230,232,234}$Ac~\cite{Bara2025} produced only upper limits for $\beta$DF probabilities of the cases studied. These results highlighted the difficulties encountered when measuring rare decay processes like $\beta$DF, as well as the need for alternative means of characterisation~\cite{Andreyev2013, Andel2025}.
	
A theoretical approach was developed with the goal of calculating $\beta$DF probabilities for all the experimentally known cases. A similar methodology was discussed in Ref.~\cite{Veselsky2012} for $^{178,180}$Tl, in Ref.~\cite{Andel2020} for $^{188}$Bi, and in Ref.~\cite{Bara2025}, where a first version of the approach presented in this work was benchmarked by calculating $\beta$DF probabilities for $^{178,180}$Tl, as well as for the neutron-rich $^{230,232}$Fr and $^{230,232,234}$Ac. The framework presented in this work is systematically benchmarked for the first time with all experimentally known $\beta$DF cases. To describe the different components needed for the calculation of $\beta$DF probabilities, energy density functional based models (EDF) were used. These are the only models that offer a microscopic description of the nuclear many-body system for heavy nuclei whilst managing to obtain good agreement with experimental observables, e.g., see Ref.~\cite{Martini2014,Andel2020} for $\beta$-decay half-lives, or Ref.~\cite{Sanchez2025, Sanchez2025_Bf} for spontaneous fission (SF) half-lives and fission barriers, respectively. 
	
The manuscript is structured as follows: Sec.~\ref{theory_framework} describes the microscopic theoretical framework developed for the calculations of $\beta$DF probabilities, beginning with the modelling of the individual components of the process ($\beta$ decay and fission), and leading to the final results obtained with the code TALYS~\cite{Koning2023}. 

In Sec.~\ref{sens_studies}, the discussion of the calculations performed is expanded through systematic studies of the dependence of the calculated probabilities on the different ingredients used. Ultimately, the framework is optimised to reproduce the experimental $\beta$DF probabilities as precisely as possible. The optimised framework is used to identify new cases of interest for future experiments (presented in Sec.~\ref{new_cases}).
	
Finally, Sec.~\ref{conclusions} summarises the theoretical findings presented in the work, and discusses the outlook of the $\beta$DF framework. In particular, the optimised configuration of the microscopic approach determined in Sec.~\ref{sens_studies} is used to find new $\beta$DF cases that could be measured experimentally based on sources that are currently available, both on the neutron-deficient and on the neutron-rich side of the nuclear chart. Lastly, points that could be improved in the theoretical framework and its future applications are discussed.

\section{Microscopic modelling of $\beta$-delayed fission}\label{theory_framework}
The $\beta$DF probability is defined as~\cite{Ghys2015}
\begin{equation}
	P_{\beta DF} = \frac{\int_{0}^{Q_\beta} S_\beta(E) f(Z, Q_\beta - E) \frac{\Gamma_f(E)}{\Gamma_\text{tot}(E)}dE}{\int_{0}^{Q_\beta} S_\beta(E) f(Z, Q_\beta - E) dE}
	\label{eq:pbdf_theory}
\end{equation}
where $Q_\beta$, Z, $S_\beta(E)$ and $f(Z, Q_\beta - E)$ are the Q-value, proton number, $\beta$-strength, and Fermi function of the mother nucleus, respectively. The quantities $E$, $\Gamma_\text{f}$, and $\Gamma_\text{tot}$ are the excitation energy, the fission decay width, and the total decay width of the daughter, respectively.

As $\beta$DF is a two-step process, distinct quantities must be modelled. First, the $\beta$ decay of the mother nucleus must be described. This translates into the calculation of the $\beta$-feeding function, as well as the determination of masses for the Q-value determination. The second stage to be characterised is the fission of the daughter, which requires determining the path that describes the evolution of the nucleus towards scission. 

At present, no single model is capable of providing all the necessary quantities outlined above simultaneously and accurately. Several models can be employed to describe the main components of $\beta$DF separately, such as the Gross theory for $\beta$ decay~\cite{Takahashi1969}, and the finite-range liquid-drop model (FRLDM) for fission~\cite{Moller2015}. However, these models provide only a macroscopic description of the nucleus\footnote{Only some microscopic corrections are included in FRLDM to account for shell and pairing effects.} and they are not directly connected to the microscopic structure of the nucleus in terms of the individual nucleons. Instead, the approach developed in this work employs different microscopic models based on energy density functionals.

In EDF models, functionals of the nuclear densities are utilised to describe the total nuclear energy~\cite{Ryssens2016}. The parameters included in the energy functionals are fitted to experimental data, carrying the risk that predictions outside the fitted domain may suffer from larger deviations. nevertheless, EDF-based models are currently the only models capable of offering a microscopic description of heavy nuclei in terms of their constituents nucleons~\cite{Ryssens2023}, thereby providing a unique tool to tackle the many-body problem. In this work, specific parametrisations of the Gogny and Skyrme interactions are used in the modelling of the $\beta$-feeding function and of the fission path, respectively. Further details will be discussed in Sec.~\ref{beta_decay} and Sec.~\ref{fission}. 

\subsection{Modelling $\beta$ decay}\label{beta_decay}
\begin{figure}
	\centering
	\includegraphics[width=\linewidth, keepaspectratio]{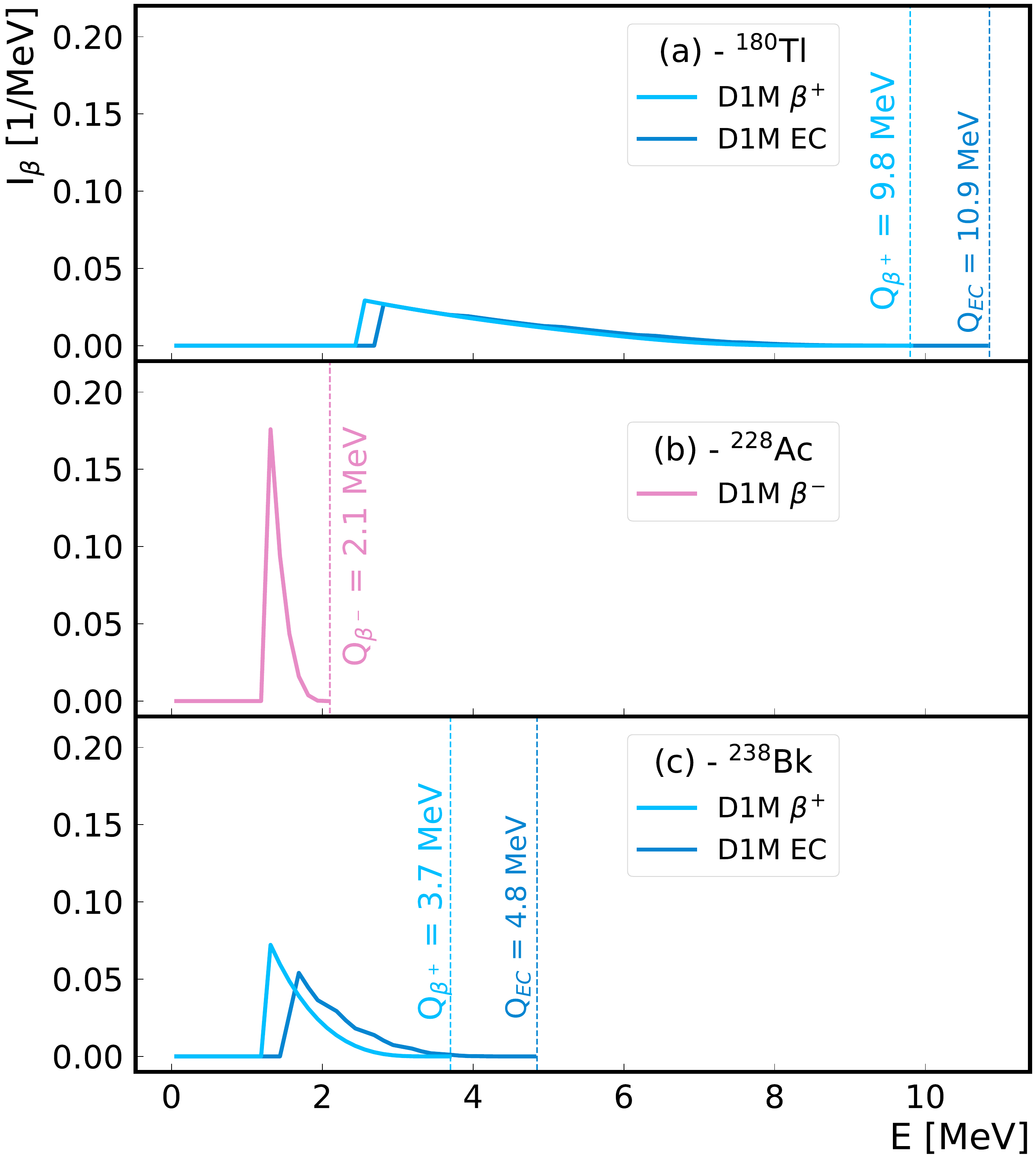}
	\caption[I$_{\beta^\pm/\text{EC}}(\text{E})$ calculated for (a) $^{180}$Tl, (b) $^{228}$Ac, and (c) $^{238}$Bk]{Plot of the I$_{\beta^\pm/\text{EC}}(\text{E})$ calculated for (a) $^{180}$Tl, (b) $^{228}$Ac, and (c) $^{238}$Bk using the approach described in Ref.~\cite{Martini2014}. The I$_\text{EC}$(E) was obtained by `stretching' the I$_{\beta^+}$ to the Q$_\text{EC}$ value for the neutron-deficient cases.}
	\label{fig:b_strength_plots}
\end{figure}

The $\beta$-decay probability strongly depends on the $\beta$-strength function, $S_\beta(E)$. This is not a straightforward quantity to calculate, as it requires knowledge of the initial and final nuclear state wave functions. 
The approach adopted for the calculations of the $\beta$-strength functions in this work is the proton-neutron quasiparticle random-phase approximation (pnQRPA) based on axially symmetric-deformed Hartree-Fock-Bogoliubov (HFB) calculations with the D1M parametrisation of the Gogny density functional~\cite{Martini2014}. The D1M+QRPA approach was chosen for its self-consistency, as the same nucleon-nucleon effective interaction is used to describe both the ground state and the excited states. Moreover, this framework offers a large-scale coverage across the nuclear chart with the largest set of calculated $\beta$-feeding functions, which is essential for the systematic application pursued in this work. The Fermi function for the $\beta$-decay is also included, so that the performed calculations yield the $\beta$-feeding to the daughter state, $\text{I}_\beta$, as an output. Only allowed Gamow-Teller transitions are included, and the $\beta$-feeding functions are calculated for $\beta^\pm$ decays. To determine $\text{Q}_\beta$ values, D1M mass predictions or experimental masses (when available) are used~\cite{Martini2014}. Figure~\ref{fig:b_strength_plots} shows the I$_{\beta^\pm}(\text{E})$ calculated for $^{180}$Tl, $^{228}$Ac, and $^{238}$Bk.

In Ref.~\cite{Martini2014}, the $\beta$-feeding functions calculated for allowed GT $\beta^-$ transitions were used to estimate the $\beta$ decay half-lives, which were subsequently compared with experimental values. Larger deviations were found for nuclei close to the valley of stability (i.e., smaller Q-values). This is a general trend for all models due to the difficulty of estimating the strength in the window of very small Q-values. The consequence is an overestimation of the experimental half-lives that, however, does not exceed one order of magnitude. A comparison of $\beta^+$ half-lives can be found in Ref.~\cite{Andel2020}.

Global alternatives to the framework used in this work include the macroscopic-microscopic FRDM+QRPA approach~\cite{Moller2003}, which offers comparably broad coverage but lacks a fully microscopic description of the underlying structure; the relativistic QRPA discussed in Ref.~\cite{Marketin2016}, which is restricted to spherical nuclei and is therefore less suited to the deformed systems studied here; and the Skyrme finite-amplitude-method QRPA~\cite{Mustonen2016, Ney2020}, which has been extended to odd-mass and odd-odd nuclei but, being developed for $\beta^-$ decay of nuclei from stability to the neutron drip line, is not immediately applicable to the nuclei in the neutron-deficient region of the chart relevant for this work. 

The $\beta$-feeding functions of the neutron-deficient cases are calculated for the $\beta^+$ decay, but these nuclei can actually undergo a combination of $\beta^+$ and EC. Therefore, the $\beta^+$ D1M $\beta$-feeding function was adapted for EC by stretching it to cover the energy range up to $Q_\text{EC}$ (see Fig.~\ref{fig:b_strength_plots}). The relative branching ratios of $\beta^+$ and EC, $\text{b}_{\beta^+}$ and $\text{b}_\text{EC}$, were calculated using a simple expression describing the ratio between the probability of EC from a K-shell ($\lambda_\text{EC,K}$) and the probability of $\beta^+$ decay ($\lambda_{\beta^+}$)~\cite{Preston}. This approximation holds for low Z and relatively high Q-values, and it is given by
\begin{equation}
	\frac{\lambda_\text{EC,K}}{\lambda_{\beta^+}} = 60\pi\cdot\left[\frac{\alpha\cdot m_ec^2\cdot(Z-1)}{Q_{\beta^+}+m_ec^2}\right]^3
	\label{eq:K_EC_bp_ratio}
\end{equation}
where $\alpha$ is the fine structure constant, $Z$ is the proton number of the mother, $Q_{\beta^+}$ is the $\beta^+$ Q-value, and $m_ec^2$ is the electron rest mass. This expression shows that for small Z and/or high Q-values, EC decay is less probable than $\beta^+$ decay. Although this approximation provides merely a coarse representation of $\text{b}_{\beta^+}$ and $\text{b}_\text{EC}$, it can nevertheless serve as a preliminary indicative estimate.

Furthermore, some of the experimentally studied cases involve a mother nucleus $\beta$-decaying from an isomeric state. However, this characteristic was not taken into account in the calculation of the $\beta$-feeding function and, as a consequence, of the $\beta$DF probability. Only ground-state (g.s.) $\beta$ decay was considered. Energy-wise, this approximation should not make a significant difference, as the isomer states involved have relatively low energies (e.g., the energy of $^{202\mathrm{m}}$Fr is 257(6)\;keV~\cite{Kondev_2021} compared to a g.s.\;to g.s.\;Q-value of about 8\;MeV). The main impact might come from the spin-parity assigned to the isomer state, as this would influence the states populated in the daughter, thus potentially changing the trend of the $\beta$-feeding function. However, the spin-parity assignment for the cases considered is not always known experimentally. Therefore, the calculations discussed below ignore the contribution of isomeric states.

\subsection{Modelling fission}\label{fission}

To model fission, a description of the nuclear evolution towards the scission point is required. The most natural coordinate to describe fission is the mass quadrupole deformation, $\beta_{20}$~\cite{Ryssens2016}. To account for the degree of reflection asymmetry as the nucleus evolves towards scission, the octupole deformation, $\beta_{30}$, must also be included. This quantity determines how different in mass the emitted fission fragments (FF) will be, and, therefore, whether the fission is symmetric or asymmetric~\cite{Ryssens2016}. Furthermore, it has been shown in several works (see e.g., Ref.~\cite{Larsson1972, Scamps2021, Ryssens2022}) that the nucleus frequently ceases to remain axially symmetric in its evolution towards fission, instead acquiring triaxial deformation, $\beta_{22}$. 

The chosen deformation parameters ($\beta_{20}$, $\beta_{22}$, and $\beta_{30}$) constitute the set of collective coordinates utilised to compute the potential energy surface (PES) for the nucleus of interest. The self-consistent EDF-based model employed in this work is BSkG3, from the Brussels-Skyrme-on-a-Grid series~\cite{Grams2023}. BSkG3 was selected for its ability to accurately reproduce SF half-lives and actinide fission barriers within a single self-consistent functional~\cite{Sanchez2025, Sanchez2025_Bf}. Gogny-force-based alternatives~\cite{Schunck2016, Bender2020} offer a comparable microscopic base and have been extensively benchmarked for fission in the actinides; however, a systematic study across the nuclear chart similar to Ref.~\cite{Sanchez2025_Bf} had not been performed at the time this work was carried out. In the broader fission study reported in Ref.~\cite{Sanchez2025_Bf}, when comparing empirical fission barriers from RIPL3~\cite{Capote2009} to those obtained with the same methodology discussed in this work, a very good agreement was found, with deviations of 300\;-\;500\;keV and never exceeding 1\;MeV. However, the comparison was performed only for nuclei with Z between 90 and 96 (which were also used for the fit of the model's parameters), therefore, different deviations might be expected in other regions.

The total energy calculated with BSkG3 is composed of a mean-field contribution from the HFB equations, which also contains the pairing energy, and a collective correction part that includes both rotational and vibrational corrections~\cite{Grams2023}. The different points of the energy surface are distributed on a grid and calculated at fixed values of the chosen collective coordinate (e.g., ($\beta_{20}$, $\beta_{30}$)) by iteratively solving the HFB equations with the Skyrme effective interaction using the MOCCa code~\cite{Ryssens2016}. A spacing of approximately $\Delta\beta\;\sim$ 0.005\;-\;0.05 is normally used. The final PES is obtained from the interpolation of the calculated points. 

To determine the fission probability, the evolution of a nucleus is followed from the g.s. to the scission point, which is identified from the particle density in the neck region~\cite{Perez2017}. Specifically, the least-action path (LAP), $L$, that minimises the action, $S(L)$, must be determined. The action along the path can be expressed as~\cite{Flynn2022}
\begin{equation}
	S(L)=\frac{1}{\hbar}\int_{s_\text{in}}^{s_\text{out}}\sqrt{2M_\text{eff}(s)\left[V_\text{eff}(s)-E_\text{GS}\right]}\;ds
	\label{eq:action}
\end{equation}
where $M_\text{eff}$ and $V_\text{eff}$ are the effective inertia and the effective potential energy along the path, respectively, and $E_\text{GS}$ is the ground state energy. The integral is carried over the points of the path $ds$ between the extremes, $s_\text{in}$ and $s_\text{out}$, which are the initial and final points of the path, respectively. The effective inertia is defined as~\cite{Flynn2022}
\begin{equation}
	M_\text{eff} = \sum_{i,j}\mathcal{M}_{i,j}(\mathbf{q})\frac{dq_i}{ds}\frac{dq_j}{ds}
	\label{eq:eff_inertia}
\end{equation}
where $\mathcal{M}_{i,j}(\mathbf{q})$ is the collective inertia (or mass) tensor derived from the adiabatic time-dependent HFB model (ATDHFB) with the cranking approximation~\cite{Baran2011}, and $\mathbf{q}$ represents the collective variables.

\begin{figure*}
	\centering
	\includegraphics[width=\textwidth, keepaspectratio]{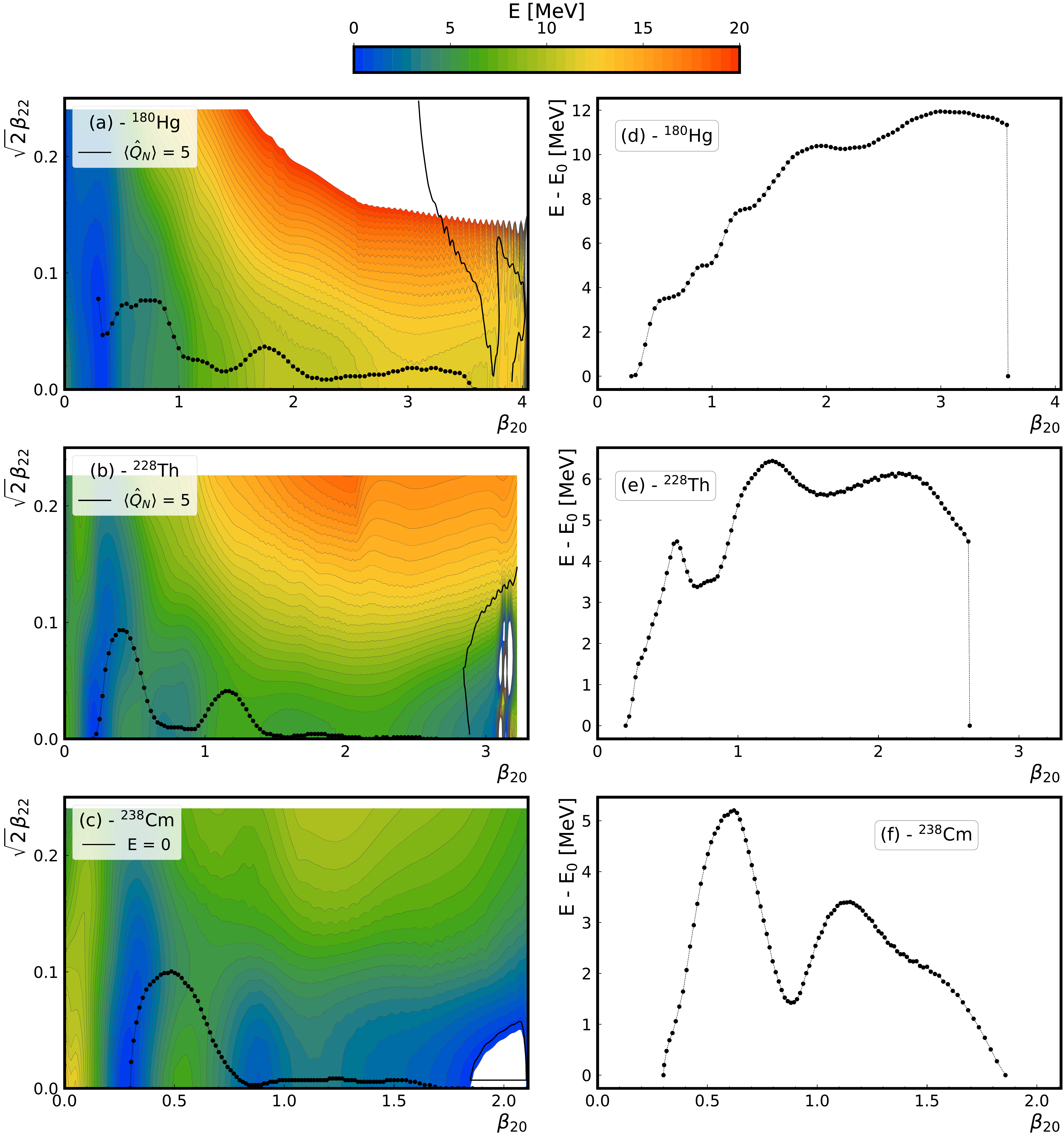}
	\caption[LAPs and 1D projections in the ($\beta_{20}$, $\beta_{22}$) coordinate space]{PES with calculated LAPs and respective 1D-projections showing the potential energy as a function of $\beta_{20}$ for (a) $^{180}$Hg, (b) $^{228}$Th, and (c) $^{238}$Cm in the ($\beta_{20}$, $\beta_{22}$) coordinate space.}
	\label{fig:LAPs}
\end{figure*}

The LAP is extracted from the calculated PES as a function of the different collective coordinates using the nudged elastic band method as implemented in PyNEB~\cite{Flynn2022}. The two-step strategy presented in~\cite{Sanchez2025} was applied: first, a least-energy path (LEP) is extracted from a ($\beta_{20}$, $\beta_{30}$) PES to constrain the octupole deformation with respect to the quadrupole deformation. The constrained octupole deformation is then used to compute the ($\beta_{20}$, $\beta_{22}$) PES, from which the final LAP is obtained. Examples of LAPs calculated in the ($\beta_{20}$, $\beta_{22}$) coordinate space, as well as 1D-projections of the nuclear energy along the fission path, are shown in Fig.~\ref{fig:LAPs}. 
Following the completion of the calculations for this work, it was discovered that PyNEB can converge to local minima of the action, mostly affecting cases where $s_\text{in}$ and $s_\text{out}$ are axial configurations. In such cases, the paths were exploring triaxiality while they should have remained axially symmetric. However, even for these nuclei, if one accounts for the initial excitation energy of the compound nucleus, the effective reduction of the barriers favours triaxial paths. For the cases analysed in this work with excitation energies up to the Q-values of the mother nuclei, the LAPs are triaxial. A publication detailing more information about the energy dependence of LAPs is currently in preparation~\cite{Sanchez2026}. 

The calculations were performed for all experimentally known $\beta$DF cases, whose details can be found in Table~\ref{tab:details_n_def} and Table~\ref{tab:details_n_rich} for the neutron-deficient and neutron-rich cases, respectively. Details of the obtained calculations will be shown only for $^{180}$Tl, $^{228}$Ac, and $^{238}$Bk. These nuclei have been chosen as examples of the three larger regions of the nuclear chart where $\beta$DF experimental values can be found: the neutron-deficient lead and transuranium regions for $^{180}$Tl and $^{238}$Bk, respectively, and the neutron-rich region below uranium for $^{228}$Ac. For more details regarding the fission path determination using PyNEB, tests on the stability of the obtained paths, and the results obtained for all experimentally known $\beta$DF cases, see Ref.~\cite{Bara_phd}.

\begin{table*}
	\centering
	\caption[Neutron-deficient cases details]{Details of the neutron-deficient cases considered: half-lives, $Q_{\beta^+}$ values of the mother nuclei~\cite{NNDC}, and primary fission barrier B$_\text{f}$ of the daughter from the BSkG3 framework~\cite{Grams2023}. The table also includes rough estimates of the $\text{b}_{\beta^+}$ and $\text{b}_\text{EC}$ calculated using Eq.~\ref{eq:K_EC_bp_ratio} for the mother nuclei.}
	\label{tab:details_n_def}
	\begin{tabular}{cccccc}
		\toprule[0.8pt]
		\midrule[0.8pt]
		\multirow{2}{*}{\textbf{Decay}} & \multirow{2}{*}{\textbf{T$_{1/2}^\text{M}$}} & \boldmath$Q_{\beta^+}$ & \boldmath$\text{b}_{\beta^+}$  &\boldmath$\text{b}_\text{EC}$ & \textbf{B$_\text{f}^\text{BSkG3}$}\\
		& & \textbf{[MeV]} &  \textbf{[\%]} & \textbf{[\%]} & \textbf{[MeV]}\\
		\midrule[0.8pt]
		\midrule[0.8pt]
		$^{178}$Au $\longrightarrow$ $^{178}$Pt &3.4(5)\;s & 8.67 &  99.40 & 0.60 & 13.61\\
		\midrule[0.8pt]
		$^{178}$Tl $\longrightarrow$ $^{178}$Hg &0.25(2)\;s & 10.68 &  99.64 & 0.36 & 10.74\\
		$^{180}$Tl $\longrightarrow$ $^{180}$Hg &0.94(25)\;s & 9.84 &  99.55 & 0.45 & 11.94\\
		\midrule[0.8pt]
		$^{186}$Bi $\longrightarrow$ $^{186}$Po &14.8(8)\;s & 10.51 &  99.60 & 0.40 & 9.66\\
		$^{188\text{ls},\text{hs}}$Bi $\longrightarrow$ $^{188}$Po &60, 265\;ms & 9.59 &  99.48 & 0.52 & 10.83\\
		\midrule[0.8pt]
		$^{192\text{m1},\text{m2}}$At $\longrightarrow$ $^{192}$Pb & 88(6), 11.5(6)\;ms & 9.97 &  99.50 & 0.50 & 7.55\\
		$^{194\text{m1},\text{m2}}$At $\longrightarrow$ $^{194}$Pb & 286(7), 323(7)\;ms & 9.27 &  99.38 & 0.62 & 8.76\\
		$^{196}$At $\longrightarrow$ $^{196}$Pb & 371(5)\;ms & 8.53 &  99.22 & 0.78 & 10.09\\
		\midrule[0.8pt]
		$^{200}$Fr $\longrightarrow$ $^{200}$Rn & 49.4(4)\;ms & 9.11 &  99.31 & 0.69 & 8.56\\
		$^{202\text{gs},\text{m}}$Fr $\longrightarrow$ $^{202}$Rn & 372(12), 286(13)\;ms & 8.35 &  99.11 & 0.89 & 10.18\\
		\midrule[0.8pt]
		$^{228}$Np $\longrightarrow$ $^{228}$U & 61.4(14)\;s & 3.58 &  89.99 & 10.01 & 5.27\\
		\midrule[0.8pt]
		$^{230}$Am $\longrightarrow$ $^{230}$Pu & 0.52\;min & 4.92 &  95.17 & 4.83 & 3.38\\
		$^{232}$Am $\longrightarrow$ $^{232}$Pu & 1.31(4)\;min & 4.00 &  91.87 & 8.13 & 4.07\\
		$^{234}$Am $\longrightarrow$ $^{234}$Pu & 2.32(8)\;min & 4.11 &  92.39 & 7.61 & 4.78\\
		\midrule[0.8pt]
		$^{236}$Bk $\longrightarrow$ $^{236}$Cm & 22$^{+13}_{-6}$\;s & 4.70 &  94.24 & 5.76 & 4.53\\
		$^{238}$Bk $\longrightarrow$ $^{238}$Cm & 144(5)\;s & 3.70 &  89.61 & 10.39 & 5.2\\
		$^{240}$Bk $\longrightarrow$ $^{240}$Cm & 4.2(8)\;min & 2.92 &  82.35 & 17.65 & 5.81\\
		$^{242}$Bk $\longrightarrow$ $^{242}$Cm & 7.0(13)\;min & 1.93 &  62.70 & 37.30 & 5.96\\
		\midrule[0.8pt]
		$^{240}$Es $\longrightarrow$ $^{240}$Cf & 6(2)\;s & 5.20 &  95.29 & 4.71 & 5.05\\
		$^{242}$Es $\longrightarrow$ $^{242}$Cf & 11(3)\;s & 4.40 &  92.79 & 7.21 & 5.68\\
		$^{244}$Es $\longrightarrow$ $^{244}$Cf & 38(11)\;s & 3.53 &  87.76 & 12.24 & 6.64\\
		$^{246}$Es $\longrightarrow$ $^{246}$Cf & 7.7(5)\;min & 2.71 &  78.40 & 21.60 & 6.38\\
		$^{248}$Es $\longrightarrow$ $^{248}$Cf & 23(3)\;min & 2.04 &  64.33 & 35.67 & 6.07\\
		\midrule[0.8pt]
		$^{250}$Md $\longrightarrow$ $^{250}$Fm & 52(6)\;s & 3.3 &  84.98 & 15.02 & 8.02\\
		\midrule[0.8pt]
		\bottomrule[0.8pt] 
	\end{tabular}
\end{table*}

\begin{table*}
	\centering
	\caption[Neutron-rich cases details]{Same as Table~\ref{tab:details_n_def} but for the neutron-rich nuclides studied.}
	\label{tab:details_n_rich}
	
	\begin{tabular}{ccccc}
		\toprule[0.8pt]
		\midrule[0.8pt]
		\multirow{2}{*}{\textbf{Decay}} & \multirow{2}{*}{\textbf{T$_{1/2}^\text{M}$}} & \boldmath$\text{Q}_{\beta^-}$ & \boldmath$\text{b}_{\beta^-}$  & \textbf{B$_\text{f}^\text{BSkG3}$}\\
		& & \textbf{[MeV]} & \textbf{[\%]} & \textbf{[MeV]}\\
		\midrule[0.8pt]
		\midrule[0.8pt]        
		$^{228}$Fr $\longrightarrow$ $^{228}$Ra & 38(1)\;s & 4.44 & 100 & 7.72\\
		$^{230}$Fr $\longrightarrow$ $^{230}$Ra & 19.1(5)\;s &  4.97 & 100 & 7.51\\
		$^{232}$Fr $\longrightarrow$ $^{232}$Ra & 5.5(6)\;s &  5.58 & 100 & 7.46\\
		\midrule[0.8pt]
		$^{228}$Ac $\longrightarrow$ $^{228}$Th & 6.15(2)\;h &  2.12 & 100 & 6.44\\
		$^{230}$Ac $\longrightarrow$ $^{230}$Th & 122.0(30)\;s &  2.98 & 100 & 6.27\\
		$^{232}$Ac $\longrightarrow$ $^{232}$Th & 119.0(50)\;s &  3.71 & 100 & 6.42\\
		$^{234}$Ac $\longrightarrow$ $^{234}$Th & 44.0(70)\;s &  4.23 & 100 & 6.59\\
		\midrule[0.8pt]
		$^{234\mathrm{gs,m}}$Pa $\longrightarrow$ $^{234}$U & 6.70(5)\;h, 1.159(11)\;min &  2.19 & 100 & 5.72\\
		$^{236}$Pa $\longrightarrow$ $^{236}$U & 9.1(1)\;min &  2.89 & 100 & 6.14\\
		$^{238}$Pa $\longrightarrow$ $^{238}$U & 2.3(1)\;min &  3.59 & 100 & 6.34\\
		\midrule[0.8pt]
		$^{256\mathrm{m}}$Es $\longrightarrow$ $^{256}$Fm & 7.6\;h & 1.70 & 100 & 5.38\\
		
		\midrule[0.8pt]
		\bottomrule[0.8pt]
	\end{tabular}
\end{table*}

\subsubsection{TALYS}\label{talys}
The final $\beta$DF probability is calculated by feeding all the different ingredients that describe $\beta$DF in TALYS~\cite{Koning2023}, a code developed to simulate nuclear reactions. For the $\beta$DF case, the target nucleus is the fissioning daughter, and the projectile is the $\beta$-decaying mother nuclide. Therefore, the inputs TALYS requires to calculate $\beta$DF probabilities are the $\beta$-feeding of the mother nucleus, alongside the fission path and nuclear level-density (NLD) of the daughter. The latter is calculated at the saddle-points identified from the fission path using the HFB plus combinatorial method described in Ref.~\cite{Goriely2008, Hilaire2006}. The $\beta$-feeding of the mother is used to determine the population of different states in the daughter. The fission paths and the nuclear level densities calculated at the saddle points are used to determine the fission transmission coefficients through each barrier: for excitation energies below the fission barrier, the WKB approximation is used, while the Hill-Wheeler expression is applied otherwise~\cite{Koning2023}. The final transmission coefficient is given by the following expression~\cite{Koning2023}
\begin{align}
	T_f(E,J,\pi) & = \sum_i T_f(E,J,\pi, \epsilon_i) +\nonumber \\
	& + \int_{E_{th}}^{E} \rho(\epsilon,J,\pi)T_f(E,J,\pi, \epsilon) d\epsilon
	\label{eq:full_trans_coeff}
\end{align}
where the summation runs over discrete energy levels, $\epsilon_i$, on top of the barrier, while the integration is described by the level density, $\rho(\epsilon,J,\pi)$, determined at each saddle point, and starts from the threshold energy, $E_{th}$, which marks the beginning of the continuum.

If the fission path presents multiple maxima, TALYS uses the so-called full damping limit to calculate the total transmission coefficient through all barriers. The full damping limit has the general form of $1/T_f = \sum_{i=1}^{N} 1/T_i$ and the transmission coefficients are calculated independently for the individual barriers.

As the final output, TALYS returns the $\beta$DF probability $\text{P}_{\beta DF}$, which is defined as
\begin{equation}
	P_{\beta DF} = \sum_{E, J, \pi} \frac{\text{I}_\beta(E,J,\pi)\cdot T_f(E,J,\pi)}{T_f(E,J,\pi) + \sum_\text{ch}T_\text{ch}(E,J,\pi)}
	\label{eq:fiss_cs}
\end{equation}
where $\text{I}_\beta(E,J,\pi)$ is the $\beta$-feeding function of the mother nucleus to the fissioning nucleus state of energy $E$, and spin-parity, $J^\pi$, while $T_f(E,J,\pi)$ is the fission transmission coefficient of the considered level and $T_\text{ch}(E,J,\pi)$ is the transmission of all competing channels (e.g., $\gamma$ decay, p-/n-emission, etc.)~\cite{Capote2009}. 

The fission path used in the TALYS calculations is extracted from the zero-temperature configuration, while the fissioning states of the daughter nuclei populated by $\beta$ decay lie at finite excitation energy up to $Q_\beta$. TALYS accounts for this only implicitly, through the excitation-energy dependence of the level density entering the transmission coefficient (Eq.~\ref{eq:full_trans_coeff}), rather than by recomputing the barrier itself at a finite temperature. Finite-temperature microscopic calculations show that both the barrier height and the PES topology can change appreciably with excitation energy~\cite{Sheikh2009, Schunck2015, Zhao2019};therefore, this static-path approach is a known approximation.

\subsection{Calculated probabilities}\label{raw_probs}

The results obtained from the full calculation of $\beta$DF probabilities are presented in the second column of Table~\ref{tab:final_calc_results_ndef} and Table~\ref{tab:final_calc_results_nrich} for the neutron-deficient and the neutron-rich nuclei, respectively. The second-to-last column reports the deviation of the calculated probabilities from the experimental values expressed in orders of magnitude, for cases with finite experimental $\text{P}_{\beta DF}$. Although neutron-deficient mother nuclei can undergo a combination of $\beta^+$ and EC decay, as a first approximation, only the $\beta^+$ decay was included. Furthermore, as anticipated in Sec.~\ref{beta_decay} cases in which the mother decays from an isomeric state were not taken into account, and all mothers are assumed to decay from the g.s.. 

\begin{table*}
	\centering
	\caption[Values of $\text{P}_{\beta DF}$ calculated for neutron-deficient nuclei]{Final $\beta$DF probabilities obtained for the neutron-deficient cases studied. The decay considered for the calculations is the $\beta^+$ decay or a combination of $\beta^+$ decay and EC decay. The columns report (starting from the left) the mother-daughter decay considered, the `raw' calculated value of $\text{P}_{\beta DF}$, the probability obtained with the optimised methodology described in Sec.~\ref{opt_results}, and the experimental values. As a first approximation, all mother nuclei are assumed to decay from the g.s.. For finite values of experimental $\text{P}_{\beta DF}$, the last two columns report the logarithm of the probabilities ratio, to show the deviations of the calculated values from the experimental ones expressed in orders of magnitude.}
	\label{tab:final_calc_results_ndef}
	\resizebox{\textwidth}{!}{%
		\begin{tabular}{ccccccc}
			\toprule[0.8pt]
			\midrule[0.8pt]
			\textbf{Decay} & \boldmath$\text{P}_{\beta DF}^\text{th}$ &\boldmath{P$_{\beta DF}^\text{th-opt}$} & \multicolumn{2}{c}{\boldmath$\text{P}_{\beta DF}^\text{exp}$} & \boldmath$\text{log}_{10}\left(\frac{\text{P}_{\beta DF}^\text{th}}{\text{P}_{\beta DF}^\text{exp}}\right)$ & \boldmath$\text{log}_{10}\left(\frac{\text{P}_{\beta DF}^\text{th-opt}}{\text{P}_{\beta DF}^\text{exp}}\right)$\\[1mm]
			\midrule[0.8pt]
			\midrule[0.8pt]        
			$^{178}$Au $\longrightarrow$ $^{178}$Pt &  1.2$\times$10$^{-18}$ &2.0$\times$10$^{-14}$ & $<$\;9.7(2)$\times$10$^{-9}$ & \cite{Andel2025} & /&/\\[1mm]
			\midrule[0.8pt]
			$^{178}$Tl $\longrightarrow$ $^{178}$Hg &  1.5$\times$10$^{-9}$ &1.0$\times$10$^{-3}$ & 1.5(6)$\times$10$^{-3}$ & \cite{Liberati2013} & -6&0\\[1mm]
			$^{180}$Tl $\longrightarrow$ $^{180}$Hg &  5.4$\times$10$^{-12}$ &2.7$\times$10$^{-7}$ & 3.2(2)$\times$10$^{-5}$ & \cite{Elseviers2013} & -7&-2\\[1mm]
			\midrule[0.8pt]
			$^{186\mathrm{gs,m}}$Bi $\longrightarrow$ $^{186}$Pb &  1.8$\times$10$^{-6}$ &9.2$\times$10$^{-3}$ & 7.6$\times$10$^{-2}$ & \cite{Lane2013} & -5&-1\\[1mm]
			$^{188\mathrm{ls}}$Bi $\longrightarrow$ $^{188}$Pb &  8.3$\times$10$^{-12}$ & 2.3$\times$10$^{-4}$ & (0.4(2)\;-\;1.8(7))$\times$10$^{-3}$ & \cite{Andel2020} & / & / \\[1mm]
			$^{188\mathrm{hs}}$Bi $\longrightarrow$ $^{188}$Pb &  8.3$\times$10$^{-12}$ & 2.3$\times$10$^{-4}$ & 4.6(9)$\times$10$^{-3}$& \cite{Andel2020} & -9& -1\\[1mm]
			\midrule[0.8pt]
			$^{192\mathrm{m1,m2}}$At $\longrightarrow$ $^{192}$Po &  2.6$\times$10$^{-3}$ & 9.4$\times$10$^{-2}$ & (0.7\;-\;3.5)$\times$10$^{-1}$ & \cite{Andreyev2013_At} & /&/\\[1mm]
			$^{194\mathrm{m1,m2}}$At $\longrightarrow$ $^{194}$Po &  4.8$\times$10$^{-7}$ & 4.5$\times$10$^{-3}$ & (0.8\;-\;1.6)$\times$10$^{-2}$ & \cite{Andreyev2013_At} & /&/\\[1mm]
			$^{196}$At $\longrightarrow$ $^{196}$Po &  8.1$\times$10$^{-12}$ & 4.6$\times$10$^{-6}$ & 9(1)$\times$10$^{-5}$ & \cite{Truesdale2016} & -7&-1\\[1mm]
			\midrule[0.8pt]
			$^{200}$Fr $\longrightarrow$ $^{200}$Rn &  5.9$\times$10$^{-5}$ & 2.3$\times$10$^{-2}$ & (0.31(17)\;-\;1.9(16))$\times$10$^{-1}$ & \cite{Ghys_thesis} & /&/\\[1mm]
			$^{202\mathrm{gs,m}}$Fr $\longrightarrow$ $^{202}$Rn &  6.9$\times$10$^{-16}$ & 1.4$\times$10$^{-5}$ & 3.0(5)$\times$10$^{-4}$ & \cite{Ghys_thesis, Ghys2015} & -12& -1\\[1mm]
			\midrule[0.8pt]        
			$^{228}$Np $\longrightarrow$ $^{228}$U &  3.1$\times$10$^{-16}$ & 1.5$\times$10$^{-8}$ & 2.0(9)$\times$10$^{-4}$ & \cite{Kreek1994} & -12&-4\\[1mm]
			\midrule[0.8pt]
			$^{230}$Am $\longrightarrow$ $^{230}$Pu &  3.4$\times$10$^{-1}$ & 8.9$\times$10$^{-1}$ & $>$\;3$\times$10$^{-1}$ & \cite{Wilson2017} & / & /\\[1mm]
			$^{232}$Am $\longrightarrow$ $^{232}$Pu &  1.0$\times$10$^{-2}$ & 1.6$\times$10$^{-2}$ & 6.9(10)$\times$10$^{-4}$ & \cite{Hall1990} & 1 & 1\\[1mm]
			$^{234}$Am $\longrightarrow$ $^{234}$Pu &  1.4$\times$10$^{-5}$ & 8.7$\times$10$^{-5}$ & 6.6(18)$\times$10$^{-5}$ & \cite{Hall1990_234am} & -1 & 0\\[1mm]
			\midrule[0.8pt]
			$^{236}$Bk $\longrightarrow$ $^{236}$Cm &  2.0$\times$10$^{-2}$ & 2.3$\times$10$^{-2}$ & 4.0(20)$\times$10$^{-2}$ & \cite{Konki2017} & 0 & 0\\[1mm]
			$^{238}$Bk $\longrightarrow$ $^{238}$Cm &  1.2$\times$10$^{-4}$ & 3.5$\times$10$^{-4}$ & 4.8(20)$\times$10$^{-4}$ & \cite{Kreek1994_2} & -1 & 0\\[1mm]
			$^{240}$Bk $\longrightarrow$ $^{240}$Cm &  1.3$\times$10$^{-7}$ & 3.1$\times$10$^{-7}$ & $\left(1.3^{+1.8}_{-0.7}\right)\times$10$^{-5}$ &\cite{Galeriu1983} & -2 & -2\\[1mm]
			$^{242}$Bk $\longrightarrow$ $^{242}$Cm &  3.2$\times$10$^{-9}$ & 8.2$\times$10$^{-6}$ & $<$\;3.0$\times$10$^{-7}$ & \cite{Gangrsky1978} & / & /\\[1mm] 
			\midrule[0.8pt]    
			$^{240}$Es $\longrightarrow$ $^{240}$Cf &  4.4$\times$10$^{-2}$ & 4.6$\times$10$^{-1}$ & 1.6(6)$\times$10$^{-1}$ & \cite{Konki2017} & -1 & 0\\[1mm]
			$^{242}$Es $\longrightarrow$ $^{242}$Cf &  6.7$\times$10$^{-5}$ & 1.9$\times$10$^{-2}$ & 4(1)$\times$10$^{-3}$ & \cite{Shaughnessy2000, Andel2025} & -2 & 1\\[1mm]
			$^{244}$Es $\longrightarrow$ $^{244}$Cf &  9.2$\times$10$^{-8}$ & 4.4$\times$10$^{-5}$ & 1.2(4)$\times$10$^{-4}$ & \cite{Shaughnessy2002} & -3 & 0\\[1mm]
			$^{246}$Es $\longrightarrow$ $^{246}$Cf &  5.1$\times$10$^{-10}$ & 1.0$\times$10$^{-6}$ & $\left(3.7^{+8.5}_{-3.0}\right)\times$10$^{-5}$ & \cite{Shaughnessy2001} & -5 & -2\\[1mm]
			$^{248}$Es $\longrightarrow$ $^{248}$Cf &  1.2$\times$10$^{-9}$ & 3.5$\times$10$^{-6}$ & 3.5(18)$\times$10$^{-6}$& \cite{Shaughnessy2001} & -3 & 0\\[1mm]
			\midrule[0.8pt]
			$^{250}$Md $\longrightarrow$ $^{250}$Fm &  7.6$\times$10$^{-9}$ & 3.8$\times$10$^{-6}$ & $\left(2^{+2}_{-1}\right)\times$10$^{-4}$ & \cite{Gangrsky1980} & -4 & -2\\[1mm]
			\midrule[0.8pt]
			\bottomrule[0.8pt] 
		\end{tabular}}
\end{table*}

\begin{table*}
	\centering
	\caption[Values of $\text{P}_{\beta DF}$ calculated for neutron-rich nuclei]{Same as Table~\ref{tab:final_calc_results_ndef} but for the neutron-rich nuclides studied. In this case the decay considered is the $\beta^-$. As a first approximation, all mother nuclei are assumed to decay from the g.s.. For finite values of experimental $\text{P}_{\beta DF}$, the last two columns report the logarithm of the probabilities ratio, to show the deviations of the calculated values from the experimental ones expressed in orders of magnitude.}
	\label{tab:final_calc_results_nrich}
	\begin{tabular}{ccccccc}
		\toprule[0.8pt]
		\midrule[0.8pt]
		\textbf{Decay} & \boldmath$\text{P}_{\beta DF}^\text{th}$ &\boldmath{P$_{\beta DF}^\text{th-opt}$} & \multicolumn{2}{c}{\boldmath$\text{P}_{\beta DF}^\text{exp}$} & \boldmath$\text{log}_{10}\left(\frac{\text{P}_{\beta DF}^\text{th}}{\text{P}_{\beta DF}^\text{exp}}\right)$ & \boldmath$\text{log}_{10}\left(\frac{\text{P}_{\beta DF}^\text{th-opt}}{\text{P}_{\beta DF}^\text{exp}}\right)$\\[1mm]
		\midrule[0.8pt]
		\midrule[0.8pt]        
		$^{228}$Fr $\longrightarrow$ $^{228}$Ra &  5.0$\times$10$^{-17}$ & 2.7$\times$10$^{-13}$ & $<$\;2.0$\times$10$^{-7}$ & \cite{Mezilev1990} & / & /\\[1mm]
		$^{230}$Fr $\longrightarrow$ $^{230}$Ra &  6.5$\times$10$^{-16}$ &  2.1$\times$10$^{-11}$ & $<$\;3.5$\times$10$^{-8}$& \cite{Bara_phd} & / & /\\[1mm]
		$^{232}$Fr $\longrightarrow$ $^{232}$Ra &  9.2$\times$10$^{-15}$ & 1.4$\times$10$^{-8}$ & $<$\;1.3$\times$10$^{-7}$ & \cite{Bara2025} & / & /\\[1mm]
		\midrule[0.8pt]
		$^{228}$Ac $\longrightarrow$ $^{228}$Th &  5.0$\times$10$^{-13}$ & 5.0$\times$10$^{-13}$ & 5.0(20)$\times$10$^{-12}$ & \cite{Yanbing2006} & -1 & -1\\[1mm]
		$^{230}$Ac $\longrightarrow$ $^{230}$Th &  1.3$\times$10$^{-11}$ & 1.3$\times$10$^{-11}$ & $<$\;4.7$\times$10$^{-10}$ & \cite{Bara_phd} & / & / \\[1mm]
		$^{232}$Ac $\longrightarrow$ $^{232}$Th &  1.7$\times$10$^{-19}$ & 1.7$\times$10$^{-19}$ & $<$\;3.2$\times$10$^{-9}$ & \cite{Bara_phd} & / & / \\[1mm]
		$^{234}$Ac $\longrightarrow$ $^{234}$Th &  3.3$\times$10$^{-11}$ & 3.3$\times$10$^{-11}$ & $<$\;4.9$\times$10$^{-6}$ & \cite{Bara2025} & / & / \\[1mm]
		\midrule[0.8pt]
		$^{234\mathrm{gs}}$Pa $\longrightarrow$ $^{234}$U &  6.3$\times$10$^{-17}$ & 6.3$\times$10$^{-17}$ & 3.0$\times$10$^{-12}$ & \cite{Gangrsky1978} & -5 & -5 \\[1mm]
		$^{234\mathrm{m}}$Pa $\longrightarrow$ $^{234}$U &  6.3$\times$10$^{-17}$ & 6.3$\times$10$^{-17}$ & 1.0$\times$10$^{-12}$ & \cite{Gangrsky1978} & -4 & -4 \\[1mm]
		$^{236}$Pa $\longrightarrow$ $^{236}$U &  8.7$\times$10$^{-14}$ & 8.7$\times$10$^{-14}$ & 1.0$\times$10$^{-9}$ & \cite{Batist1977} & -4 & -4 \\[1mm]
		$^{238}$Pa $\longrightarrow$ $^{238}$U &  5.2$\times$10$^{-11}$ & 5.2$\times$10$^{-11}$ & $<$\;2.6$\times$10$^{-8}$  & \cite{Baas-May1985} & / & / \\[1mm]
		\midrule[0.8pt]
		$^{256\mathrm{m}}$Es $\longrightarrow$ $^{256}$Fm &  9.3$\times$10$^{-8}$ & 3.4$\times$10$^{-5}$ & 2.0$\times$10$^{-5}$  & \cite{Hall1989} & -2 & 0 \\[1mm]
		
		\midrule[0.8pt]
		\bottomrule[0.8pt]
	\end{tabular}
\end{table*}

\section{Sensitivity studies}\label{sens_studies}
Various tests have been performed to study the influence of the quantities used in the calculations on the final results. The parameters that were found to have the largest effect are the scaling of the energy along the fission path and the $\beta$-feeding function used, which are discussed in Sec.~\ref{en_fission_path} and Sec.~\ref{b_feeding}, respectively. Other quantities varied include the level density and the scaling of the inertia; however, either no effect or no global trend across the different nuclei studied was found.

Optimising the quantities used based on the calculation for a single nucleus would not be practical when expanding the calculations discussed in this work to a broader group of nuclei, such as those of interest for r-process nucleosynthesis. However, treating all nuclei in the same way might be a too rough an approximation. Therefore, the goal of the tests discussed below is to find a global configuration that describes the $\beta$DF process as precisely as possible throughout the different regions of the nuclear chart. Although the sensitivity studies have been performed for all $\beta$DF cases that have a finite value of $\text{P}_{\beta DF}$, only the cases of $^{180}$Tl, $^{238}$Bk, and $^{228}$Ac will be explicitly shown below. 

To quantify how well the calculated probabilities reproduce the experimental values, the root-mean-square (RMS) deviation can be defined as
\begin{equation}
	\delta_\text{RMS} = \sqrt{\frac{1}{N}\sum_{i=0}^{N}\left(\text{log}_{10}\frac{P_{\beta DF}^\text{th}}{P_{\beta DF}^\text{exp}}\right)^2}
	\label{eq:prob_RMS}
\end{equation}
where $N$ is the number of cases considered, and $P_{\beta DF}^\text{th}$ and $P_{\beta DF}^\text{exp}$ are the calculated and experimental values of the probabilities, respectively. With this definition, $\delta_\text{RMS}$ quantifies the orders of magnitude by which the calculations globally deviate from the experimental values. To calculate $\delta_\text{RMS}$ only the experimental cases that have finite $\text{P}_{\beta DF}$ values will be considered, and only qualitative considerations will be made for the cases where experimental limits are reported.

\subsection{Energy along the fission path}\label{en_fission_path}
\begin{figure*}
	\centering
	\includegraphics[width=0.8\linewidth, keepaspectratio]{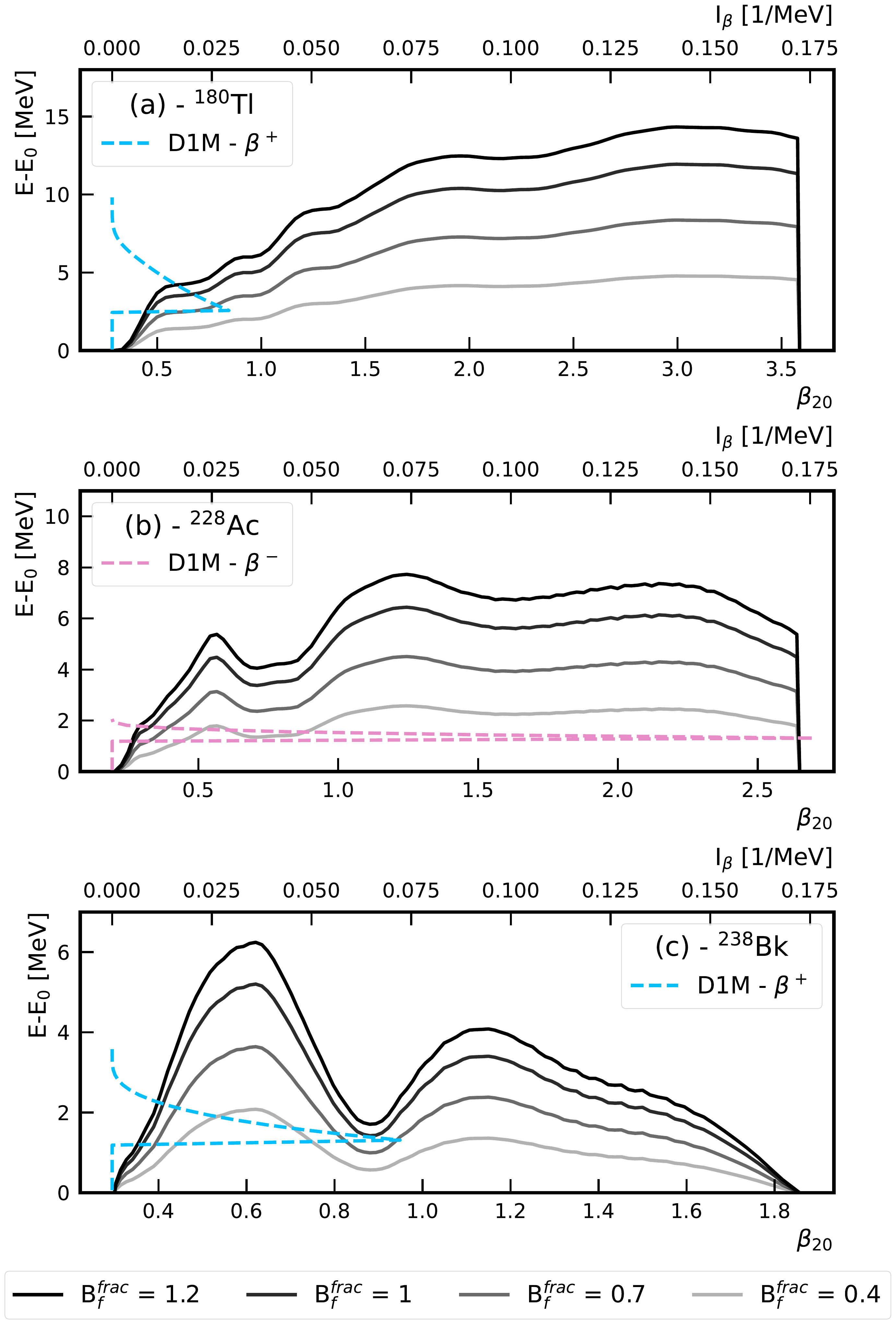}
	\caption[1D paths at different $\text{B}_\text{f}^\text{frac}$]{Energy along the fission path as a function of $\beta_{20}$ for (a) $^{180}$Hg, (b) $^{228}$Th, and (c)  $^{238}$Cm, scaled with different $\text{B}_\text{f}^\text{frac}$ values. The $\beta$-feeding function of the mother nucleus ((a) $^{180}$Tl, (b) $^{228}$Ac, and (c) $^{238}$Bk) calculated with the D1M approach is plotted as well for comparison with the energy spanned by the fission path. Note the different scales of the x-axes and of the y-axes.}
	\label{fig:1D_paths_vs_bstrength_morevfiscor}
\end{figure*}

The first sensitivity study investigates the influence of the energy along the fission path on the final $\beta$DF probability. As the fission transmission coefficient is exponentially related to the fission barrier, an overall scaling of the energy along the fission path has a strong impact on the final probability. A constant offset could also be applied, but it would not change the height of the barrier. Therefore, a linear scaling is the simplest variation of the energy along the fission path to be applied. This can be implemented directly through TALYS, and Fig.~\ref{fig:1D_paths_vs_bstrength_morevfiscor} shows the energy along the fission path obtained with scaling values of 0.4, 0.7, and 1.2, compared to the unscaled energy. The scaling factor is indicated as $\text{B}_\text{f}^\text{frac}$ to identify the fraction of the energy along the fission path (identified as the primary fission barrier, $\text{B}_\text{f}$, for simplicity) considered for the calculations. In the same figures, the $\beta$-feeding function used for the calculations is plotted to show how the energy range covered by $\text{I}_\beta$ becomes closer to the primary barrier height as the energy along the fission path is scaled further.

For $^{234,236,238}$Pa, the fission parameters used with TALYS for their daughters, $^{234,236,238}$U, could be adjusted to those needed to reproduce neutron-induced fission cross-sections~\cite{RIPL4}, while for all other cases, a range of scaling factors between 0.4 and 1.2 was tested. This is a conservative sensitivity test, as primary fission barriers are normally reproduced within a few hundred of keV by BSkG3, as mentioned in Sec.~\ref{fission}.

The results of the energy scaling test are shown in Fig.~\ref{fig:Bf_test}. The $\beta$-feeding function used is that for $\beta^\pm$ decay calculated with the D1M approach. From Fig.~\ref{fig:Bf_test} it can be observed that at very small scaling factors, the $\beta$DF probability approaches 1. This is because, in certain cases, the energy along the fission path is reduced so much that most of the energy range covered by the Q-value of the mother starts to be close to and above the primary barrier, making fission highly favourable. However, as discussed above, such small scaling factor values are not realistic, and should not be used to derive strong conclusions. 

While the unscaled energy along the fission paths already provides a good agreement with the experimental $\text{P}_{\beta DF}$ values of $^{228}$Ac and $^{238}$Bk, the calculations for $^{180}$Tl approach the experimental probability at $\text{B}_\text{f}^\text{frac}$ values between 0.7 and 0.8. Although the three cases discussed here are only examples, a similar situation is found for all others considered in this work. In particular, it is notable that unscaled energies along the fission paths yield good agreement with experimental values for cases where the daughter has a proton number between 90 and 96, while for the others, the energy along the fission path should be decreased by a factor of $\sim$\;20\%. A reduction of the energy along the fission path by 20\% corresponds to a variation of a few hundred keV from the barrier obtained from the PyNEB calculations, in line with what was observed in Ref.~\cite{Sanchez2025_Bf}. In that study, BSkG3 was found to overestimate the barriers of pre-actinides by about 10\;-\;20\% and up to 40\% for lighter nuclei such as $^{196}$Hg. Moreover, the independent study in Ref.~\cite{Andel2020} highlighted the need for a similar barrier reduction (a factor of 0.72) to reproduce the measured $\beta$DF partial half-life of $^{188hs}$Bi using a D1M-Gogny fission path. The findings in these different studies are, thus, consistent with the adjustment adopted in this work both in magnitude and in direction of the scaling.

The optimal configuration to reproduce the experimental values can be identified by examining the lowest RMS deviation for the different situations. Table~\ref{tab:RMS_Bf_Sb_tot} reports the RMS calculated with Eq.~\ref{eq:prob_RMS} for the cases considered. For the neutron-rich cases, using a different scaling factor does not significantly change the $\delta_\text{RMS}$ as there is only one case that falls in the region where the $\text{B}_\text{f}^\text{frac}$ used would be different, i.e., $^{256}$Es. On the other hand, for the neutron-deficient cases (and, as a consequence, for all nuclei considered together) adjusting the scaling of the energy along the fission path reduces the RMS deviation by a factor $\sim$\;2. 

\subsection{Influence of the $\beta$-feeding function}\label{b_feeding}
The other main component of the $\text{P}_{\beta DF}$ calculations is the $\beta$-feeding function. Only limited information on $\text{I}_\beta$ calculated with other formalisms for all the required cases is available in the literature (see, for example, the SkO' functions used in Ref.~\cite{Bara2025}). Therefore, the various $\beta$-feeding functions used were all derived from the D1M parametrisation, adjusting it for EC decay in neutron-deficient nuclei.

\begin{figure*}
	\centering
	\includegraphics[width=0.8\linewidth, keepaspectratio]{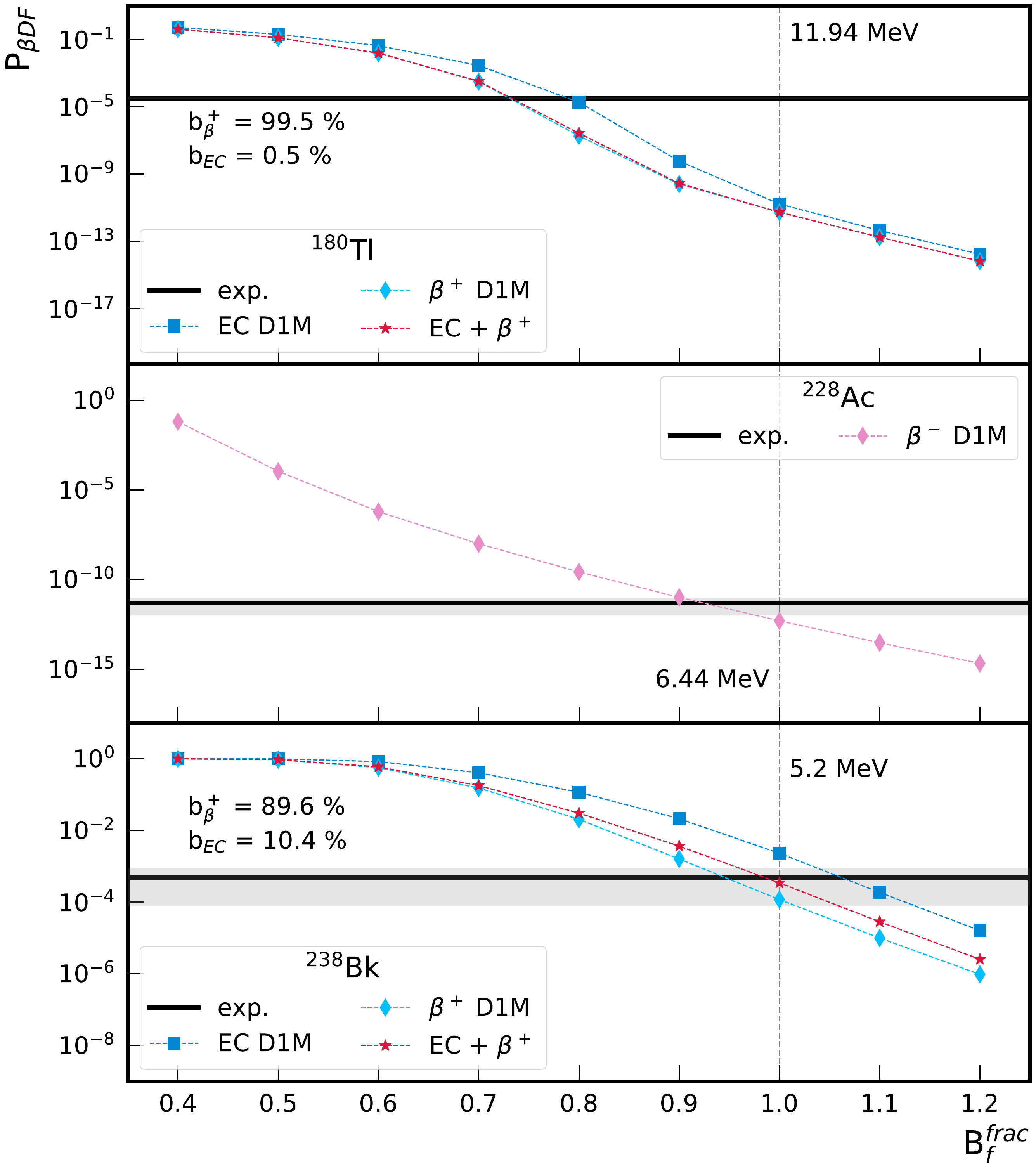}
	\caption[$\text{B}_\text{f}^\text{frac}$ + $\beta$-feeding test]{Sensitivity of the $\beta$DF probability on the scaling factor of the energy along the fission path used. The black horizontal line represents the experimental value and the grey band its uncertainty. The vertical dashed line corresponds to the probability obtained from the unscaled energy along the fission path (the value of the primary barrier is indicated next to the vertical line). 
	Different $\beta$-feeding functions have been tested, including the combination of $\beta^+$ and EC, obtained using the roughly estimated branching ratios, $\text{b}_{\beta^+}$ and $\text{b}_{EC}$, shown in the plots (only for the neutron-deficient cases).}
	\label{fig:Bf_test}
\end{figure*}

The sensitivity studies over different scaling factors of the energy along the fission paths were repeated with $\text{I}_\text{EC}$, and the results are shown in Fig.~\ref{fig:Bf_test}. The final $\beta$DF probability was obtained as a weighted average of the probabilities calculated with pure $\beta^+$- and EC-feeding functions with the formula
\begin{equation}
	P_{\beta DF}^{\text{EC}+\beta^+} = \frac{\text{b}_{\beta^+}\cdot P_{\beta DF}^{\beta^+}+\text{b}_\text{EC}\cdot P_{\beta DF}^\text{EC}}{100}
	\label{eq:p_bDF_EC_bp}
\end{equation}
The values of $b_{\beta^+/\text{EC}}$ are given in percent and correspond to the relative branching ratios of $\beta^+$ and EC, which were calculated using the expression shown in Eq.~\ref{eq:K_EC_bp_ratio}. Although Table~\ref{tab:details_n_def} reports the branching ratios for all the neutron-deficient nuclei studied, only those having finite values of $\beta$DF probabilities are used for the calculations in this sensitivity test. The results obtained using the EC-feeding function, as well as combining it with the $\beta^+$ decay in the test with the different scaling of the energy along the fission path, are shown in Fig.~\ref{fig:Bf_test}(a) and (c). While for $^{180}$Tl combining the probability obtained from pure $\beta^+$ and EC decays does not significantly alter the result obtained from the $\beta^+$ decay only, for $^{238}$Bk the situation is slightly different. The $\text{b}_\text{EC}$ of $^{238}$Bk is not negligible in the way it is for $^{180}$Tl, and therefore, including the EC contribution to the total $\beta$DF probability helps to better reproduce the experimental value. 		

Once again, to quantify the effect of using a different $\beta$-feeding function, the RMS deviation from experimental values can be calculated. Table~\ref{tab:RMS_Bf_Sb_tot} reports the $\delta_\text{RMS}$ obtained using Eq.~\ref{eq:prob_RMS} for different combinations of $\beta$-feeding function and scaling of the energy along the fission path studied.

In all combinations shown in Table~\ref{tab:RMS_Bf_Sb_tot}, the smallest deviations in the neutron-deficient group are found for $\text{B}_\text{f}^\text{frac}$\;=\;0.8 applied for nuclei whose daughter has $Z_d$\;$<$\;90 and $Z_d$\;$>$\;96, while $\text{B}_\text{f}^\text{frac}$\;=\;1 is used for the others, as discussed in the previous section. The results are further improved if EC is taken into account. As shown in Table~\ref{tab:details_n_def}, for the nuclei in the lead region of the nuclear chart (where $\text{Q}_{\beta^+}$ is relatively high), the contribution of EC to the total $\beta$ decay is negligible, while for the others, it becomes increasingly significant as the mass of the nuclei increases. Thus, considering EC alone would not be a realistic scenario for most of the nuclei studied, and the contribution of both $\beta^+$ and EC should be accounted for.

\begin{table*}
	\centering
	\caption[RMS for $\text{B}_\text{f}$ + $\text{I}_\beta$ test]{RMS deviation calculated with Eq.~\ref{eq:prob_RMS} for neutron-deficient, neutron-rich, and all cases together using different combinations of $\beta$-feeding functions and scaling of energy along the fission paths. The $\beta$-feeding function used for the neutron-deficient and the neutron-rich nuclei when considering all cases together are reported in \textcolor{deepskyblue}{light blue} and \textcolor{nrichpink}{pink}, respectively. The value N in parentheses indicates the number of cases included in each group.}
	\label{tab:RMS_Bf_Sb_tot}
	\begin{tabular}{w{c}{0.2cm}|w{c}{4.2cm}|w{c}{1.5cm}w{c}{1.5cm}w{c}{2cm}}
		\hline
		\hline 
		\multirow{9}{*}{\rotatebox[origin=c]{90}{\textbf{neutron-deficient (N\;=\;18)}}} & \multirow{2}{*}{\diagbox[height=1.05cm, innerwidth=4.2cm, innerleftsep=0.9cm, innerrightsep=0.9cm]{\boldmath$\text{B}_\text{f}^\text{frac}$}{\boldmath$\text{I}_\beta$}}
		& \makebox{\textbf{D1M} }&\makebox{\textbf{D1M}}&\makebox{\textbf{D1M}} \\
		& & \makebox{\boldmath$\beta^+$} & \makebox{\textbf{EC}} &\makebox{\textbf{EC+\boldmath$\beta^+$}} \\
		\cmidrule[0.8pt]{2-5}
		\cmidrule[0.8pt]{2-5}
		& 1 for all& 5.7 & 3.7 & 4.6 \\
		& 0.9 for $Z_d$\;$<$\;90 $\wedge$ $Z_d$\;$>$\;96 & \multirow{2}{*}{4.1} & \multirow{2}{*}{2.3} & \multirow{2}{*}{2.7} \\
		& 1 for 90\;$<$\;$Z_d$\;$<$\;96 & & &\\
		& 0.8 for $Z_d$\;$<$\;90 $\wedge$ $Z_d$\;$>$\;96 & \multirow{2}{*}{3.1} & \multirow{2}{*}{1.1} & \multirow{2}{*}{1.8}\\
		& 1 for 90\;$<$\;$Z_d$\;$<$\;96 & & & \\
		& 0.7 for $Z_d$\;$<$\;90 $\wedge$ $Z_d$\;$>$\;96 & \multirow{2}{*}{2.9} & \multirow{2}{*}{1.5} & \multirow{2}{*}{2.4}\\
		& 1 for 90\;$<$\;$Z_d$\;$<$\;96 & & &\\
		\hline
		\hline
		\multirow{8}{*}{\rotatebox[origin=c]{90}{\textbf{neutron-rich (N\;=\;5)}}} & \diagbox[innerwidth=4.2cm, innerleftsep=0.9cm, innerrightsep=0.9cm]{\boldmath$\text{B}_\text{f}^\text{frac}$}{\boldmath$\text{I}_\beta$} & \multicolumn{3}{c}{\makebox{\textbf{D1M} \boldmath$\beta^-$}} \\
		\cmidrule[0.8pt]{2-5}
		\cmidrule[0.8pt]{2-5}
		& 1 for all& \multicolumn{3}{c}{3.5} \\
		& 0.9 for $Z_d$\;$<$\;90 $\wedge$ $Z_d$\;$>$\;96 & \multicolumn{3}{c}{\multirow{2}{*}{3.4}} \\
		& 1 for 90\;$<$\;$Z_d$\;$<$\;96 & & & \\
		& 0.8 for $Z_d$\;$<$\;90 $\wedge$ $Z_d$\;$>$\;96 & \multicolumn{3}{c}{\multirow{2}{*}{3.4}} \\
		& 1 for 90\;$<$\;$Z_d$\;$<$\;96 & & & \\
		& 0.7 for $Z_d$\;$<$\;90 $\wedge$ $Z_d$\;$>$\;96 & \multicolumn{3}{c}{\multirow{2}{*}{3.5}} \\
		& 1 for 90\;$<$\;$Z_d$\;$<$\;96 & & & \\
		\hline
		\hline
		\multirow{9}{*}{\rotatebox[origin=c]{90}{\textbf{all nuclei (N\;=\;23)}}} & \multirow{2}{*}{\diagbox[height=1.05cm, innerwidth=4.2cm, innerleftsep=0.9cm, innerrightsep=0.9cm]{\boldmath$\text{B}_\text{f}^\text{frac}$}{\boldmath$\text{I}_\beta$}}
		& \makebox{\textbf{D1M} }&\makebox{\textbf{D1M}}&\makebox{\textbf{D1M}} \\
		& & \makebox{\textcolor{deepskyblue}{\boldmath$\beta^+$}/\textcolor{nrichpink}{\boldmath$\beta^-$}} & \makebox{\textcolor{deepskyblue}{\textbf{EC}}/\textcolor{nrichpink}{\boldmath$\beta^-$}} & \makebox{\textcolor{deepskyblue}{\textbf{EC+\boldmath$\beta^+$}}/\textcolor{nrichpink}{\boldmath$\beta^-$}}\\
		\cmidrule[0.8pt]{2-5}
		\cmidrule[0.8pt]{2-5}
		& 1 for all& 5.3 & 3.6 & 4.4 \\
		& 0.9 for $Z_d$\;$<$\;90 $\wedge$ $Z_d$\;$>$\;96 & \multirow{2}{*}{3.9} & \multirow{2}{*}{2.6} & \multirow{2}{*}{2.9}\\
		& 1 for 90\;$<$\;$Z_d$\;$<$\;96 & & &\\
		& 0.8 for $Z_d$\;$<$\;90 $\wedge$ $Z_d$\;$>$\;96 & \multirow{2}{*}{3.1} & \multirow{2}{*}{1.9} & \multirow{2}{*}{2.0} \\
		& 1 for 90\;$<$\;$Z_d$\;$<$\;96 & & & \\
		& 0.7 for $Z_d$\;$<$\;90 $\wedge$ $Z_d$\;$>$\;96 & \multirow{2}{*}{3.1} & \multirow{2}{*}{2.1} & \multirow{2}{*}{2.0} \\
		& 1 for 90\;$<$\;$Z_d$\;$<$\;96 & & & \\
		\hline
		\hline
	\end{tabular}
\end{table*}

Finding the optimal combination of parameters is intricate, and physical arguments should guide the choice. Scaling the energy along the fission path of some nuclei is supported by the precision of the model in the reproduction of primary fission barriers. The $\text{B}_\text{f}^\text{frac}$\;=\;0.8 choice for $Z_d$\;$<$\;90 and $Z_d$\;$>$\;96 is motivated by the accuracy of BSkG3 in reproducing barriers, as discussed in Sec.~\ref{fission}. The same reasoning can be applied to the inclusion of EC for the neutron-deficient cases, and both the $\beta^+$ and EC channels should be considered to achieve a more realistic description of the process. Therefore, calculating the final probability as a weighted average of the two obtained from pure $\beta^+$/EC-decay represents the most accurate approach. The $\beta^+$/EC branching ratios calculated with Eq.~\ref{eq:K_EC_bp_ratio} represent a rough estimate and might deviate significantly from the real true values. However, any combination of $\text{b}_{\beta^+}$ and $\text{b}_\text{EC}$ would produce a $\text{P}_{\beta DF}$ value within the probabilities calculated with pure $\beta^+$- and pure EC-feeding functions, which are about one to two orders of magnitude apart (see Fig.~\ref{fig:Bf_test}(a)-(c)). Therefore, the crude approximation used to determine $\text{b}_{\beta^+}$ and $\text{b}_\text{EC}$ is not expected to substantially influence the derived conclusions. 

\begin{figure*}
	\centering
	\includegraphics[width=\linewidth, keepaspectratio]{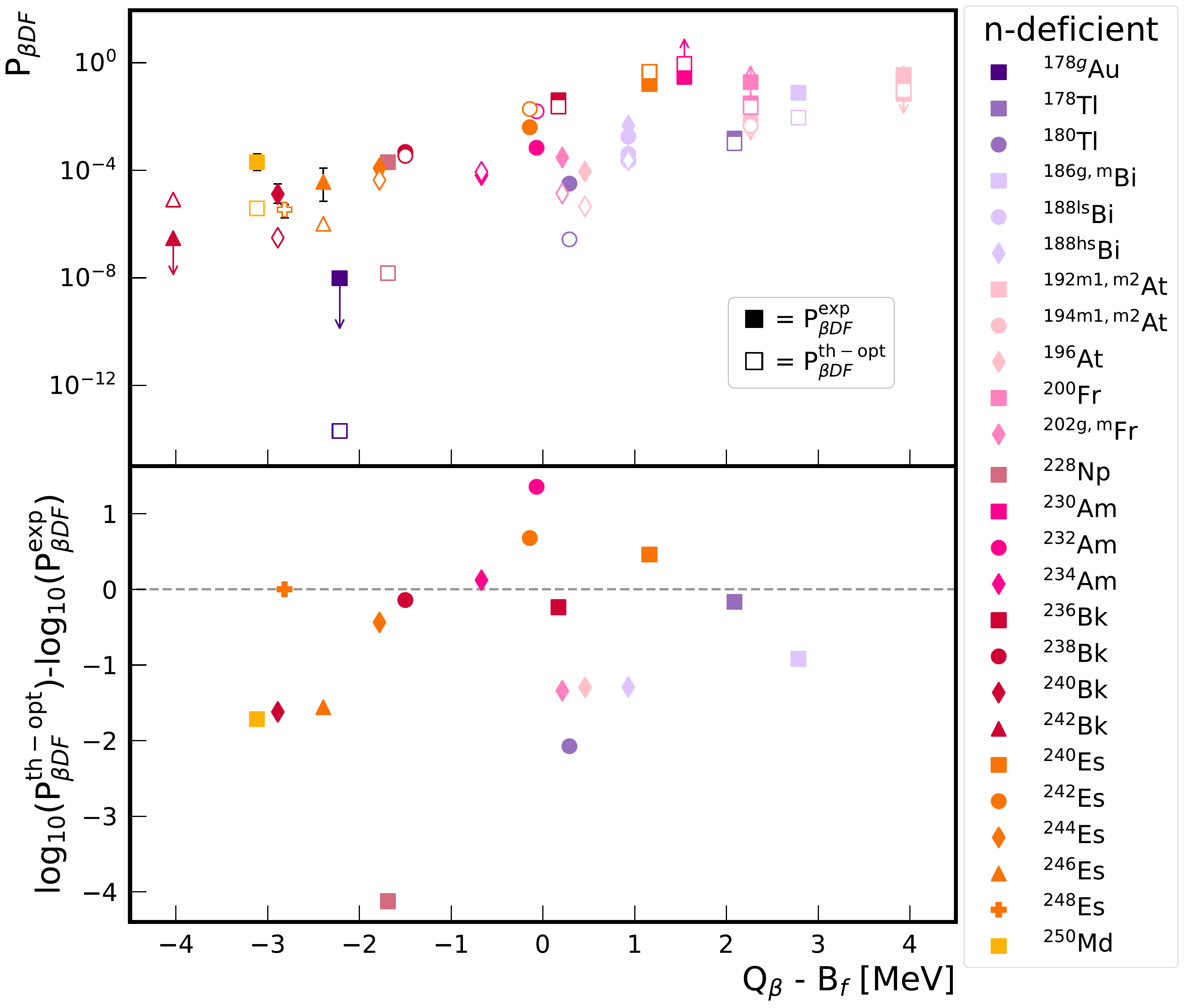}
	\caption[Comparison between experimental and final calculations (n-deficient)]{The top panel shows the comparison between the experimental $\text{P}_{\beta DF}$ (full markers) and the final calculated values (open markers) for all the neutron-deficient cases considered with the optimised microscopic approach discussed above. The bottom panel shows the deviation of the calculated value from the experimental ones of the cases with a finite $\text{P}_{\beta DF}$.}
	\label{fig:final_results_ndef}
\end{figure*}

\begin{figure*}
	\centering
	\includegraphics[width=\linewidth, keepaspectratio]{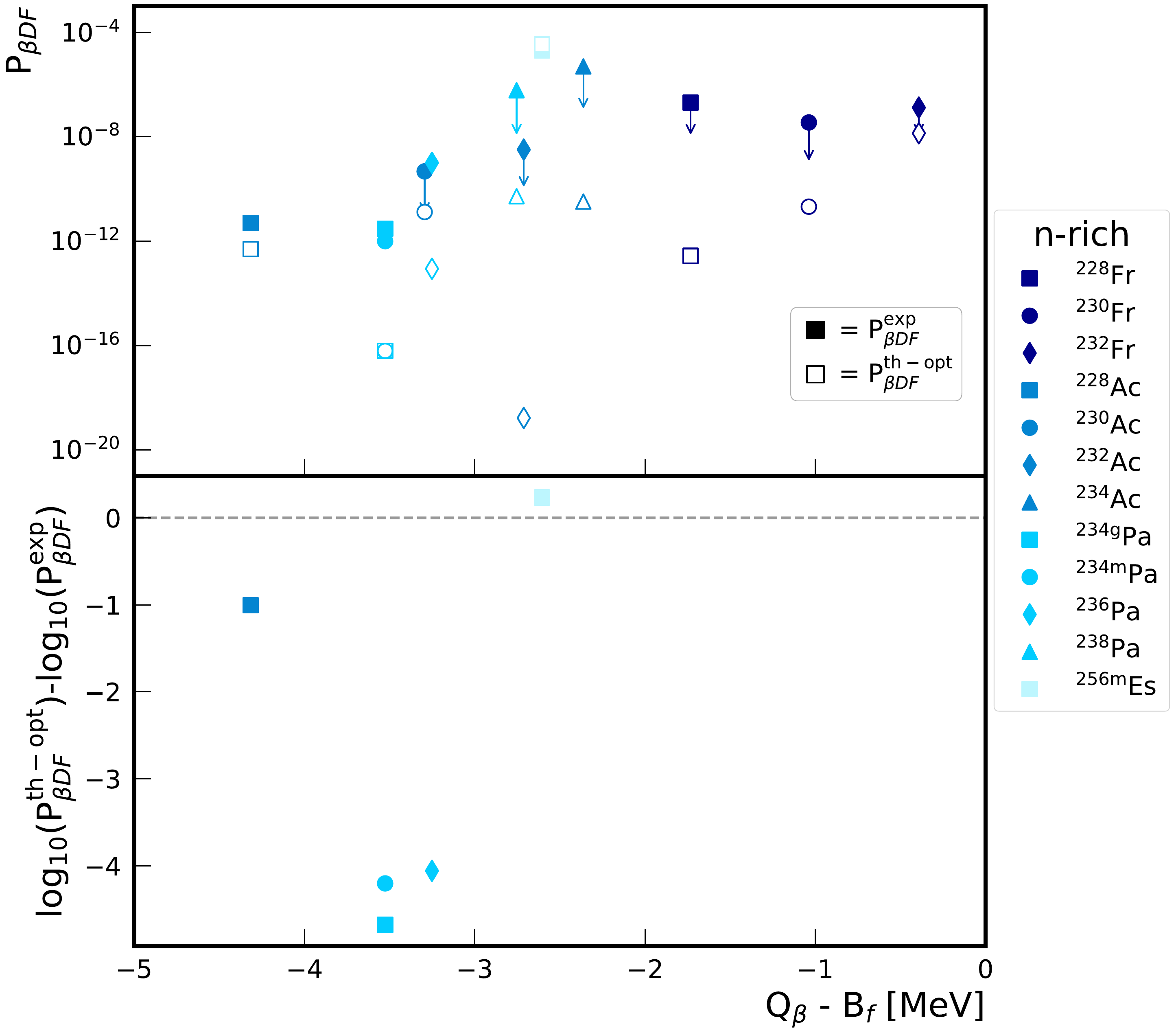}
	\caption[Comparison between experimental and final calculations (n-rich)]{Same as Fig.~\ref{fig:final_results_ndef}, but for the neutron-rich cases.}
	\label{fig:final_results_nrich}
\end{figure*}

\subsection{Optimised results}\label{opt_results}
The optimised recipe for the $\text{P}_{\beta DF}$ calculations comprises the following steps:
\begin{enumerate}
	\item Utilise the D1M $\beta$-feeding function for all cases, and include the contribution of both $\beta^+$ and EC decay channels for neutron-deficient nuclei;
	\item The presence of isomeric states in the mother nucleus is not taken into account;
	\item Reduce the energy along the fission paths of the daughters by 20\% for nuclei with $Z_d$\;$<$\;90 and $Z_d$\;$>$\;96, and leave it unscaled for the others. This step is not applicable to the $^{234,236,238}$Pa cases, where optimal scaling factor values for the daughters could be found from neutron-induced fission cross-sections.
\end{enumerate}

The fourth column of Tab.~\ref{tab:final_calc_results_ndef} and Tab.~\ref{tab:final_calc_results_nrich} shows the probabilities calculated for all neutron-deficient and neutron-rich nuclei considered, respectively, including those cases for which $\text{P}_{\beta DF}$ limits are reported. This provides at least a qualitative idea of the agreement between the experimental limits and the calculated values. The results are also shown in Fig.~\ref{fig:final_results_ndef} and Fig.~\ref{fig:final_results_nrich} for the neutron-deficient and the neutron-rich cases, respectively. The bottom panel of each figure shows the deviation of the calculated probabilities from the experimental ones expressed in orders of magnitude, for those cases having a finite value of $\text{P}_{\beta DF}$.

Most of the calculations for the cases where an experimental upper limit is reported agree with the experimental limit. In particular, for the low-spin state of $^{188}$Bi, the combination of the two isomeric states of $^{192}$At, and $^{200}$Fr, where both upper and lower limits are given, the calculation returns a value that is within the experimental range. However, for the two isomer states of $^{194}$At, the calculated $\text{P}_{\beta DF}$ does not agree with the range found experimentally. Finally, the calculated probability for $^{242}$Bk is higher than the upper limit known experimentally, even though it remains close to it.

The predictive power of the methodology discussed in this work can be compared to that of another approach used in other references, such as Ref.~\cite{Andel2025} or Ref.~\cite{LOI216_proposal}, based on $\beta$DF systematics. In particular, the systematics approach exploits an approximate exponential relationship that links the P$_{\beta DF}$ to the energy difference between the $Q_\beta$ value of the mother and the B$_\text{f}$ of the daughter. This approach is based on a fit of experimentally known values, and on a consequent extrapolation to new cases (see, e.g., Ref.~\cite{LOI216_proposal}). However, the extrapolations obtained can turn out to be several orders of magnitude away from the true probabilities. An example is the case of the neutron-rich nuclei of actinium studied experimentally in Ref.~\cite{Bara2025}. The $\beta$DF probabilities of $^{232,234}$Ac predicted with the systematics approach in Ref.~\cite{LOI216_proposal} were found to be 2-3 orders of magnitude larger than the upper limits measured in Ref.~\cite{Bara2025} confirming the poor reliability of such an approach. On the other hand, the probabilities calculated for the same cases with the optimised microscopic methodology described in this work are consistent with the experimental upper limits, suggesting a more solid predictive power of the approach.

\section{Identification of new cases}\label{new_cases}
The values reported in Table~\ref{tab:RMS_Bf_Sb_tot} suggest a global RMS deviation of approximately two orders of magnitude of the optimised microscopic approach for the `all nuclei' group. This value also offers a good starting point for experimental purposes. The developed framework can be used to calculate $\text{P}_{\beta DF}$ for $\beta$DF cases that are experimentally unknown and that could be studied in the future. The deviation of two orders of magnitude can be used as an uncertainty margin for the calculated probability to determine the conditions of a future experiment and its feasibility.

At present, appropriate beams might not be available in existing facilities for all the identified cases presented below. However, the cases considered should be plausible enough to allow target/beam development teams to find a way to produce the suggested species in amounts that would allow for the observation the $\beta$DF events. As a reference, the calculated probabilities are used to estimate the ideal intensities of the beam that should be used to make the experiments feasible within a reasonable timeframe. For all estimations performed below, a FF efficiency, $\epsilon_\text{FF}$, of 50\% was considered. This should offer a reasonable approximation, being close to the 42(3)\% determined for the experiment discussed in Ref.~\cite{Bara2025}. 

Table~\ref{tab:new_cases} reports the probabilities calculated for the suggested cases, including both neutron-deficient and neutron-rich nuclei.  The probabilities calculated with the microscopic approach were used to determine the number of ions that must be collected in order to observe a certain number of FF. The value of $\text{N}_\text{ions}$ can be determined using the following expression
\begin{equation}
	\text{N}_\text{ions} = \frac{\text{N}_\text{FF}}{\epsilon_\text{FF}\cdot \text{P}_{\beta DF} \cdot b_\beta}
	\label{eq:n_ions}
\end{equation}
where $\text{N}_\text{FF}$ is the number of FF that is desired to be observed, $\epsilon_\text{FF}$ is the FF detection efficiency, $\text{P}_{\beta DF}$ is the calculated probability, and $\text{b}_\beta$ is the $\beta$-branching ratio of the mother nuclide. The calculations below are performed assuming $\text{N}_\text{FF}$\;=\;10. This number of FF is arbitrarily chosen, but should suffice to determine the $\text{P}_{\beta DF}$ experimentally, as in most of these experiments, the background of spurious events in the energy range of the fission products is either negligible~\cite{Bara2025} or can be removed by using appropriate detection techniques. 

The number of ions depends on the implantation time and on the intensity of the implanted beam. The values of $\text{N}_\text{ions}$ can be used to determine the yield required to observe the selected statistics within a certain implantation time, e.g., the duration of a typical shift (i.e., 8\;h). Table~\ref{tab:rates_estimates} shows the yields calculated for the cases suggested.

\begin{table*}
	\centering
	\caption[Calculated $P_{\beta DF}$ for new cases]{Calculated $\beta DF$ probabilities for new neutron-deficient (a) and neutron-rich (b) cases. The columns report (starting from the left) the mother-daughter decay considered, the mother half-life, the $\beta^+$ and EC or $\beta^-$ Q-values, the $\text{B}_\text{f}$ of the daughter, the total $\beta$-branching ratio, the $\beta^+$- and EC-branching ratios roughly estimated with Eq.~\ref{eq:K_EC_bp_ratio} (only in (a)), the $P_{\beta DF}$ obtained for single $\beta^+$ and EC decays (only in (a)), and the combined value. Only g.s.\;are considered. The Q-values, half-lives and total $\beta$-branching ratio are taken from Ref.~\cite{NNDC}, while the fission barriers are extracted from the fission paths calculated with BSkG3.}
	\label{tab:new_cases}
	\subcaption*{(a)}
	\resizebox{\textwidth}{!}{%
		\begin{tabular}{cccccccccc}
			\toprule[0.8pt]
			\midrule[0.8pt]
			\multirow{2}{*}{\textbf{Decay}} & \multirow{2}{*}{\boldmath$\text{T}_{1/2}^\text{M}$} & \boldmath$Q_{\beta^+}$ &  \boldmath$\text{B}_\text{f}^\text{BSkG3}$ & \boldmath$b_{\text{EC}+\beta^+}$ & \boldmath$b_{\beta^+}$ & \boldmath$b_\text{EC}$ & \multirow{2}{*}{\boldmath$P_{\beta DF(\beta^+)}^\text{th-opt}$} & \multirow{2}{*}{\boldmath$P_{\beta DF\text{(EC)}}^\text{th-opt}$} & \multirow{2}{*}{\boldmath$P_{\beta DF\text{(tot)}}^\text{th-opt}$} \\[1mm]
			& & \textbf{[MeV]}  &  \textbf{[MeV]} & \textbf{[\%]}  & \textbf{[\%]} & \textbf{[\%]} & & & \\[1mm]
			\midrule[0.8pt]
			\midrule[0.8pt]        
			$^{204}$Fr $\longrightarrow$ $^{204}$Rn & 1.8(3)\;s & 7.56 &  11.36 & 8 & 98.8 & 1.2 & 2.3$\times$10$^{-11}$ & 2.4$\times$10$^{-9}$ & 5.1$\times$10$^{-11}$ \\[1mm]
			\midrule[0.8pt]
			$^{236}$Am $\longrightarrow$ $^{236}$Pu & 3.6(2)\;min & 2.12 &  5.24 & 100 & 69.1 & 30.9 & 1.0$\times$10$^{-6}$ & 3.9$\times$10$^{-5}$ & 1.3$\times$10$^{-5}$ \\[1mm]
			$^{238}$Am $\longrightarrow$ $^{238}$Pu & 97.4(15)\;min & 1.24 &  5.82 & 100 & 39.8 & 60.2 & 4.0$\times$10$^{-9}$ & 2.0$\times$10$^{-7}$ & 1.2$\times$10$^{-7}$ \\[1mm]
			\midrule[0.8pt]
			\bottomrule[0.8pt]
	\end{tabular}}
	\bigskip
	\subcaption*{(b)}
	\begin{tabular}{cccccc}
		\toprule[0.8pt]
		\midrule[0.8pt]
		\multirow{2}{*}{\textbf{Decay}} & \multirow{2}{*}{\boldmath$\text{T}_{1/2}^\text{M}$} & \boldmath$Q_{\beta^-}$  & \boldmath$\text{B}_\text{f}^\text{BSkG3}$ & \boldmath$b_{\beta^-}$ & \multirow{2}{*}{\boldmath$P_{\beta DF}^\text{th-opt}$} \\[1mm]
		& & \textbf{[MeV]}  & \textbf{[MeV]} & \textbf{[\%]} & \\[1mm]
		\midrule[0.8pt]
		\midrule[0.8pt]        
		$^{238}$Np $\longrightarrow$ $^{238}$Pu & 2.0993(12)\;d & 1.291 & 5.82 & 100 & 1.1$\times$10$^{-8}$ \\[1mm]
		$^{240}$Np $\longrightarrow$ $^{240}$Pu & 61.9(2)\;min & 2.191 & 5.90 & 100 & 1.1$\times$10$^{-8}$ \\[1mm]
		\midrule[0.8pt]
		$^{250}$Bk $\longrightarrow$ $^{250}$Cf & 3.212(5)\;h & 1.78 & 9.01 & 100 & 1.0$\times$10$^{-6}$ \\[1mm]
		\midrule[0.8pt]
		\bottomrule[0.8pt]
	\end{tabular}
\end{table*}

Radioactive ion beams for some of the cases presented are already available at various facilities. For example, in Ref.~\cite{Au2023} a beam intensity of 1.3(1)$\times$10$^{6}$\;pps is reported for $^{238}$Np produced from a UC$_x$ target irradiated with 1.4-GeV protons at ISOLDE. This beam intensity is two orders of magnitude above the necessary yield calculated for this case and shown in Table~\ref{tab:rates_estimates}. As a result, the experiment should remain feasible even if the calculated $\text{P}_{\beta DF}$ were a few orders of magnitude lower than the actual value, making $^{238}$Np one of the most suitable candidates to study $\beta$DF in the neutron-rich region of the nuclear chart.

Moreover, $^{238}$Am could be measured at S$^3$-LEB (GANIL, France)~\cite{Dechery2016}, which could provide a beam with an intensity of 4.4$\times$10$^5$\;pps~\cite{S3_yields}, while $^{204}$Fr could be studied at the Facility for Rare Isotope Beams, FRIB (Michigan State University, MSU)~\cite{FRIB_yields} with a radioactive ion beam of 5.2$\times$10$^7$\;pps~\cite{FRIB_yields}.

Finally, some of the cases presented could be studied using alternative methodologies. For example, the case of $^{250}$Bk could be measured by populating the nucleus of interest via the $\alpha$ decay of an offline source of $^{254}$Es (T$_{1/2}$\;=\;275.7(5)\;d~\cite{NNDC}). Due to its long half-life, $^{254}$Es would offer an optimal way to populate $^{250}$Bk for weeks, allowing for longer measurements than is typically feasible in radioactive ion beam facilities. The study of $^{238}$Np could be performed by activating $^{237}$Np in a reactor, producing an activity that would facilitate a subsequent week-long measurement.

\begin{table*}
	\centering
	\caption[Estimated required yields for new $\beta$DF experiments]{Estimated yields for new $\beta$DF experiments for the newly identified cases. The columns report the number of ions, $\text{N}_\text{ions}$, needed to observe 10 FF, assuming a $\epsilon_\text{FF}$ of 50\%. The yields are calculated assuming a measurement time, ($\text{t}_\text{meas}$), of one shift, i.e., 8\;h.} 
	\label{tab:rates_estimates}
	\begin{tabular}{cccc}
		\toprule[0.8pt]
		\midrule[0.8pt]
		\textbf{Decay} & \boldmath$\text{P}_{\beta DF\text{(tot)}}^\text{th-opt}$ & \boldmath$\text{N}_\text{ions}$ & \textbf{Yield [pps]} \\[1mm]
		\midrule[0.8pt]
		\midrule[0.8pt]        
		$^{204}$Fr $\longrightarrow$ $^{204}$Rn & 5.1$\times$10$^{-11}$ & 4.9$\times$10$^{12}$ & 1.7$\times$10$^{8}$ \\[1mm]
		\midrule[0.8pt]
		$^{236}$Am $\longrightarrow$ $^{236}$Pu & 1.3$\times$10$^{-5}$ & 1.6$\times$10$^{6}$ & 5.4$\times$10$^{1}$ \\[1mm]
		$^{238}$Am $\longrightarrow$ $^{238}$Pu & 1.2$\times$10$^{-7}$ & 1.7$\times$10$^{8}$ & 5.8$\times$10$^{3}$ \\[1mm]
		\midrule[0.8pt]        
		$^{238}$Np $\longrightarrow$ $^{238}$Pu & 1.1$\times$10$^{-8}$ & 1.9$\times$10$^{9}$ & 6.4$\times$10$^{4}$ \\[1mm]
		$^{240}$Np $\longrightarrow$ $^{240}$Pu & 1.1$\times$10$^{-8}$ & 1.8$\times$10$^{9}$ & 6.4$\times$10$^{4}$  \\[1mm]
		\midrule[0.8pt]
		$^{250}$Bk $\longrightarrow$ $^{250}$Cf & 1.0$\times$10$^{-6}$ & 9.9$\times$10$^{6}$ & 3.4$\times$10$^{2}$ \\[1mm]
		\midrule[0.8pt]
		\bottomrule[0.8pt]
	\end{tabular}
\end{table*}

\section{Conclusions}\label{conclusions}

When attempting to experimentally study $\beta$DF, difficulties related to the rarity of the process are frequently encountered. To address how to describe and characterise $\beta$DF across the nuclear landscape, a theoretical approach was developed. This framework aimed to reproduce experimentally known $P_{\beta DF}$ in both the neutron-deficient and neutron-rich regions of the nuclear chart with the highest possible precision.

The various ingredients required to calculate $\beta$DF probabilities were obtained from EDF-based models, although a single model could not be used to determine all necessary inputs simultaneously. The $\beta$-feeding function was obtained from the D1M+QRPA framework~\cite{Martini2014}, while the PES and effective inertia were based on BSkG3~\cite{Grams2023}. The fission path was extracted from the PES using the code PyNEB~\cite{Flynn2022}, which allows the calculation of LAPs in different sets of coordinates. The nuclear level density of the daughter nucleus was subsequently calculated at the saddle points determined from the fission path. Finally, all the individual ingredients were fed into TALYS.

Following sensitivity tests to study the effects of the various inputs on the calculated results, an optimal configuration for the theoretical framework was proposed. The calculations performed with this optimised and refined approach exhibit a global RMS deviation of approximately two orders of magnitude. The theoretical framework developed in this work offers a microscopic and realistic description of $\beta$DF; being refined only through a global optimisation, it should provide more reliable extrapolations compared to those obtained via the systematics approach. Although some ingredients were adjusted to data, such as the scaling of the energy along the fission path and the inclusion of EC decay, no empirical fit to the experimental data was performed. The constructed framework was therefore used to predict candidates for future experimental studies, which could be performed at various facilities to further test the precision of the calculated probabilities and broaden the general knowledge of $\beta$DF.

The framework presented in this work attempted, for the first time, to globally characterise $\beta$DF across the nuclear chart. For this initial implementation, the framework adopts certain approximations, such as pure g.s. $\beta$ decay. Several improvements could be applied: (i) including the possibility for the mother nuclei to undergo $\beta$ decay from isomeric states, e.g., following the treatment discussed in Ref.~\cite{Andel2020} for $^{188}$Bi; (ii) using a more realistic EC-feeding function or, at least, more accurate values of b$_{\beta^+/\text{EC}}$; and (iii) testing the partial damping limit instead of the full damping limit used in the calculation of the fission transmission coefficients with TALYS~\cite{Sin2006, Sin2008}. Furthermore, it would be beneficial to move beyond the single fission path description. The LAP is the most probable fission path the nucleus will follow while evolving towards scission, but it is not the only one; for example, stochastic Langevin dynamics sampling a group of trajectories around the LAP have already been used to describe FF mass and charge distributions~\cite{Sadhukhan2016}. Particularly when dealing with induced fission, it may be useful to determine the probability with which the nucleus follows different paths and how this probability is influenced by the excitation energy of the fissioning nucleus. 

Moreover, transitioning from a two-dimensional description of the nuclear PES to (at least) a three-dimensional description would be highly advantageous. Although this would increase computational costs, it would offer the possibility of exploiting an extra degree of freedom, such as the hexadecapole deformation, $\beta_{40}$, whose inclusion has been shown to remove multi-solution discontinuities and improve the PES description at large deformations~\cite{Chi2023}. Finally, having a single model capable of providing all necessary inputs would be beneficial to the overall accuracy of the framework. This would yield a consistent data set for TALYS, thereby avoiding potential uncertainties arising from the mixing of different models. 

As a final remark regarding future applications: in addition to serving as a prediction tool for new $\beta$DF cases to be measured experimentally, the calculations of $\text{P}_{\beta DF}$ should be extended to the nuclei involved in the r-process by including $\beta$DF as an available decay channel. The nuclei for which $\beta$DF has currently been studied experimentally are located far away from the region of the nuclear chart relevant to the r-process. Consequently, they cannot offer direct insights on the effects of $\beta$DF on the r-process. However, by extending the theoretical framework developed in this work to the nuclei involved in the r-process, it will become possible to determine the extent to which $\beta$DF influences its termination and, consequently, the redistribution of abundances. 

\backmatter

\bmhead{Acknowledgements}
The authors would like to acknowledge Ulli Koester for the help and interesting discussion in coming up with ideas on how to study experimentally some the new $\beta$DF cases identified.

This work has received funding from FWO (Belgium) and F.R.S.-FNRS (Belgium) under the Excellence Of Science (EOS) program (Grant No. 30468642 and No. 40007501), from the Research Foundation Flanders under Projects I002619N, I002919N, and I012420N of the International Research Infrastructure, from the KU
Leuven BOF (C14/22/104), from the FWO fellowship for fundamental research (Contract No. 1167324N). 


\bibliography{allpapers.bib}


\end{document}